\documentclass[aps,prb, twocolumn,amsmath,amssymb,longbibliography]{revtex4-1}
\usepackage{amsmath,amssymb,amsfonts,bm}
\usepackage{graphicx}
\usepackage{xcolor}
\usepackage{bbold}
\usepackage{graphicx}
\usepackage{dcolumn}
\usepackage{epstopdf}
\usepackage{epsfig}
\usepackage{url}
\usepackage{hyperref}
\usepackage{verbatim}
\usepackage[export]{adjustbox}
\usepackage{scalerel}
\usepackage{braket}
\newcommand{\Tr}{\operatorname{Tr}}

\newcommand{\be}{\begin{equation}}
\newcommand{\ee}{\end{equation}}
\newcommand{\ba}{\begin{aligned}}
\newcommand{\ea}{\end{aligned}}

\newcommand{\tthg}{t^{(\alpha)}_\mathrm{Th}}

\def\tth{t_{\rm Th}}

\DeclareMathOperator\arctanh{arctanh}

\def\ffun{f}

\def\av{{\bf a}}
\def\bv{{\bf b}}

\def\TT{W}
\def\VV{V}

\def\lmax{\lambda_{>}}

\newcommand{\thetaint}{n}
\newcommand{\citehere}[1]{[\onlinecite{#1}]}

\graphicspath{{Figures/}}

\begin{document}

\title{Spectral Lyapunov exponents in chaotic and localized many-body quantum systems}
\author{Amos Chan}
\affiliation{Princeton Center for Theoretical Science, Princeton University, Princeton NJ 08544, USA}
\author{Andrea De Luca}
\affiliation{CNRS, Universit\'{e} de Cergy-Pontoise, France
}
\author{J. T. Chalker}
\affiliation{Theoretical Physics, Oxford University, Parks Road, Oxford OX1 3PU, United Kingdom}

\date{\today}

\begin{abstract}
We consider the spectral statistics of the Floquet operator for disordered, periodically driven spin chains in their quantum chaotic and many-body localized phases (MBL). The spectral statistics are characterized by the traces of powers $t$ of the Floquet operator, and our approach hinges on the fact that, for integer $t$ in systems with local interactions, these traces can be re-expressed in terms of products of dual transfer matrices, each representing a spatial slice of the system. We focus on properties of the dual transfer matrix products as represented by a spectrum of Lyapunov exponents, which we call \textit{spectral Lyapunov exponents}. In particular, we examine the features of this spectrum that distinguish chaotic and MBL phases. The transfer matrices can be block-diagonalized using time-translation symmetry, and so the spectral Lyapunov exponents are classified according to a momentum in the time direction. For large $t$ we argue that the leading Lyapunov exponents in each momentum sector tend to zero in the chaotic phase, while they remain finite in the MBL phase. These conclusions are based on results from three complementary types of calculation. We find exact results for the chaotic phase by considering a Floquet random quantum circuit with on-site Hilbert space dimension $q$ in the large-$q$ limit. In the MBL phase, we show that the spectral Lyapunov exponents remain finite by systematically analyzing models of non-interacting systems, weakly coupled systems, and local integrals of motion. Numerically, we  compute the Lyapunov exponents for a Floquet random quantum circuit and for the kicked Ising model in the two phases. As an additional result, we calculate exactly the higher point spectral form factors (hpSFF) in the large-$q$ limit, and show that the generalized Thouless time scales logarithmically in system size for all hpSFF in the large-$q$ chaotic phase.

\end{abstract}

\maketitle
\tableofcontents

\section{Introduction} \label{intro}

\begin{figure*}[ht]
\includegraphics[width=0.95\textwidth]{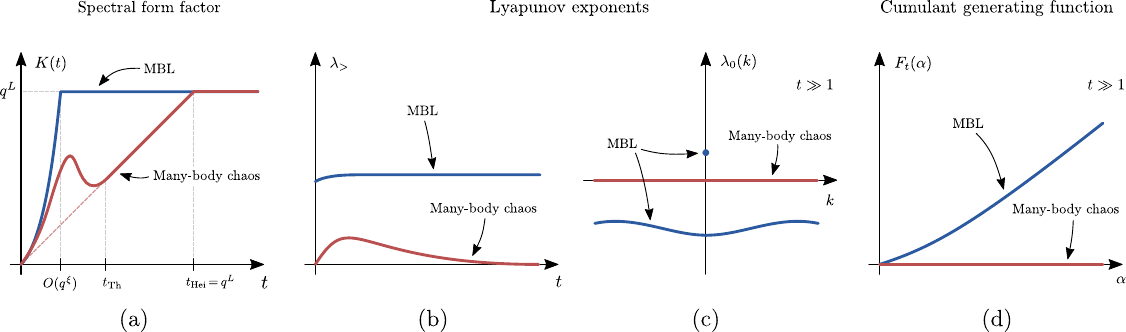}
\caption{
Summary of results: Behaviour is indicated using red lines for the quantum chaotic phase and blue lines for the MBL phase. 
(a): Schematic dependence of the SFF on $t$ for many-body chaotic and MBL Floquet systems with system size $L$.
The dashed red line is the RMT circular unitary ensemble (CUE) behavior. 
The generic behavior for many-body chaotic systems is characterized by two time scales: (i) the Thouless time $t_\mathrm{Th}$, which marks the onset of RMT behaviour in the SFF; (ii) the Heisenberg time $t_\mathrm{Hei}$, which is of the order of the inverse of mean level spacing and so scales exponentially with $L$. For MBL systems with localization length $\xi$, the SFF grows quickly with $t$, reaching a plateau at a time ${\cal O}(q^\xi)$ that is independent of $L$.
(b): Schematic dependence of the leading Lyapunov exponent $\lmax $ [see Eqns.~\eqref{eq:lyap_finiteL} and ~\eqref{eq:lambda_inf}] on $t$ in the two phases. At large $t$, $\lmax $ tends to zero in the chaotic phase and to a finite value in the MBL phase.
(c): Schematic dependence at late times of the leading Lyapunov exponent $\lambda_0(k)$ in each momentum sector on the momentum eigenvalue $k$ [see Eq.~\eqref{eq:lyap_k}]. In the chaotic phase $\lambda_0(k)$ converges to zero for all $k$. In the MBL phase $\lambda_0(k)$ converges to a smooth function of $k$ for $k\not=0$, with a distinct, larger value at $k=0$. 
(d): Schematic dependence at late times of the scaled cumulant generating function $F_t(\alpha)$ on $\alpha$ [see in Eq.~\eqref{eq:Fdef}]. The gradient of $F_t(\alpha)$ near $\alpha=0$ captures the $L$-dependence of fluctuations in the SFF. We find that $F_t(\alpha)$ has zero gradient in the chaotic phase and finite gradient in the MBL phase, reflecting fluctuations of the SFF that grow rapidly with $L$ in the second case.  
\label{fig:summary}}
\end{figure*}

One of the fundamental goals of quantum statistical mechanics is to understand the basic hallmarks of chaotic dynamics. From a practical perspective, the presence of chaos is associated with memoryless evolution, so that the thermodynamic description is well-justified on the basis of the ergodic hypothesis. However, the classical notion of chaos does not extend directly to the quantum world, as Schrodinger evolution is linear and unitary and cannot admit diverging trajectories in Hilbert space~\cite{Haake}. Nevertheless, a large class of interacting many-body systems are believed to show quantum chaotic behaviour, as embodied in the eigenstate thermalization hypothesis (ETH)\cite{Deutsch, Srednicki,Rigol2008}. By contrast, many-body localization provides a generic mechanism which prevents the onset of chaos in quantum systems in the presence of strong disorder~\cite{gornyi2005interacting, basko2006metal, MBLreview1}.


Random matrices~\cite{mehta} have long played a key role in providing minimal prototypes for properties of quantum chaotic systems. One important outcome is that spectral correlations have been identified as an indicator of chaotic behavior: as originally conjectured by Bohigas, Giannoni and Schmidt~\cite{bohigas1984characterization}, chaotic quantum systems exhibit the same spectral correlations as those of random matrices in the appropriate symmetry class. In particular, a distinctive fingerprint of quantum chaos is the presence of level repulsion between energy eigenvalues~\cite{Prosen_1993, PhysRevB.47.14291}. 

Spectral fluctuations can be conveniently characterized via the Fourier transform of the two-point correlator of eigenvalues, known as the spectral form factor (SFF): 
\begin{equation}
\label{SFFdef}
\mathcal{K}(t) =  
\sum_{m,n} 
 e^{\imath (\theta_m - \theta_n)t } = |\Tr[W(t)]|^2
 \;.
\end{equation}
Here $W$ is the generator of the time evolution and $W(t)$ denotes its $t$-power, while $\{ \theta_m \}$ are the spectral levels of the system under consideration (energies for systems with a time-independent Hamiltonian, or eigenphases of the Floquet operator for a periodically driven system).
The analysis of the SFF in many-body systems has recently been spurred on by the development of two novel approaches to Floquet models, where $W$ generates the time evolution for a single period. Both in a long-range version of the kicked Ising model,~\cite{prosen, flack2020statistics} and in Floquet random circuits~\cite{cdc1, cdc2, friedman2019, moudgalya2020spectral} in the limit of large local Hilbert space dimension, the average SFF $K(t) \equiv \langle \mathcal{K}(t) \rangle $ (where $\langle \ldots \rangle$ denotes the average over an ensemble of statistically similar systems) was shown to reproduce the RMT result for times $t$ larger than a scale $\tth$ known as the Thouless time (see Fig.~\ref{fig:summary}a). 

It has been reported on the basis of analytical and numerical calculations that $\tth$ diverges with the system size $L$ in generic quantum systems~\cite{cdc1, cdc2, friedman2019, moudgalya2020spectral, garratt2020manybody}, with the exception of specific fine-tuned models in the absence of conservation laws~\cite{bertini2018exact}. For this reason, it is important to understand which features control the behavior of $\mathcal{K}(t)$ for intermediate times $1 \ll t \lesssim \tth$, which can nevertheless be arbitrarily large in the thermodynamic limit. 
A simple argument suggests that, in this time regime, $\mathcal{K}(t)$ is typically exponentially large in $L$: because of locality of interactions, different portions of the system for $t \lesssim \tth$ have not had time to generate correlations of their eigenphases; as a consequence, the trace in \eqref{SFFdef} can be factorized into contributions from the Hilbert space of each decoupled region~\cite{cdc2}. 
Moreover, in this regime, since  the SFF is not self-averaging~\cite{PhysRevLett.78.2280} and for many-body systems has exponentially large fluctuations in the system size,
its average may not be  sufficient to characterize its behavior. 

In this paper, we study signatures of spectral statistics of quantum many-body systems with local interactions by using the fact that ${\rm Tr} [W(t)]$ can be expressed as a product of dual transfer matrices, each associated with a spatial slice of the system. The dual transfer matrix product is characterized by a set of Lyapunov exponents, which we dub the \textit{spectral Lyapunov exponents}, and by an associated cumulant generating function. The dual transfer matrix product grows exponentially with system size: average growth rates are given by Lyapunov exponents and sample-to-sample fluctuations in growth rate are described by the cumulant generating function.
There are several motivations for this approach. Knowledge of the Lyapunov exponents allows one to investigate both the spectral statistics of quantum many-body systems in the thermodynamic limit and spectral statistics at times earlier than $\tth$. In addition, knowledge of the cumulant generating function allows one to study fluctuations of the SFF.
Finally, the study spectral Lyapunov exponents provide a different way of characterising localized systems already in the thermodynamic limit.

	A summary of our results is as follows. At fixed time $t$, the spectrum of Lyapunov exponents can be organized into $t$ momentum sectors, associated with the invariance under discrete time translations of the evolution operator. We characterize the behavior of the leading Lyapunov exponent in each sector, showing that there is a clear distinction at large time ruled by the ergodicity properties of the dynamics (see Fig.~\ref{fig:summary}b and c): For chaotic systems, the largest Lyapunov exponent at each momentum sector converges to zero at large time, signaling the absence of exponential growth of $K(t)$ with system size and the emergence of random matrix behavior in the spectral correlations. For many-body localized systems, the Lyapunov exponents remain non-zero at large time, with a limiting but non-universal form of their spectrum. 
	We also discuss the fluctuations of the leading Lyapunov in the zero-momentum sector (see Fig.~\ref{fig:summary}d) by introducing a (scaled) cumulant generating function. We argue that higher cumulants are not important except in some non-generic settings.
	These results are justified by considering two different models and a combination of analytical and numerical analyses. 

	As a side result, in the chaotic phase, we compute exactly the higher point spectral form factors (hpSFF) in the limit of large local Hilbert space dimension $q\to\infty$ and thermodynamic limits. 
	The hpSFF is closely related to other diagnostics of chaos. As an example, the out-of-time-order correlator\cite{LarkinOvchinnikov, MSS, Nahum2017a, vonKeyserlingk2017, vonKeyserlingk2017a, Huse2017}  is known to be related to the hpSFF for local operators at late times\cite{cotler_decoupling_2019}, and for global operators\cite{complexity2017,van_zyl_sff_otoc_2019}. 
	We also define the generalized Thouless times as the time after which hpSFF  behaviour of a quantum many-body system reduces to the RMT result.	We show that the generalized Thouless times derived from all hpSFF scale logarithmically in system size in the large $q$ limit.  

Our calculations complement earlier work that has been concerned with the averaged SFF and its relation to chaos and localization  \cite{cdc2,Guhr2019, 2019arXiv190506345S}. In Ref.~\citehere{Guhr2019}, the growth rate of the ensemble-averaged SFF at fixed time was studied specifically for the kicked Ising model (see Sec.~\ref{subsec:KIM}) across the many-body localization transition.
The authors introduce an appropriate ensemble-averaged transfer matrix, study its symmetries, and discuss the role of the time-momentum operator. More recently a general picture was presented in Ref.~\citehere{garratt2020manybody} for the long-time behaviour of the ensemble-averaged transfer matrix, together with numerical results for a random quantum circuit in the ergodic phase. Behaviour of the ensemble-averaged transfer matrix across the MBL transition is discussed in Ref.~\citehere{garratt2020MBL}.
In contrast to this previous work, our focus here is on the average of the \textit{log} of SFF rather than of the SFF itself, and on the notion of spectral Lyapunov exponents and fluctuations in the growth rate of the dual transfer matrix product.

The remainder of this paper is organized as follows. 
In Sec.~\ref{sec:def} we introduce the spectral Lyapunov exponents and cumulant generating function. In Sec.~\ref{sec:models} we define two quantum circuit models which each display both a quantum chaotic phase and an MBL phase as a coupling parameter is varied. 
In Sec.~\ref{sec:eth_phase}, we compute exactly the Lyapunov exponents and the generating function for a random circuit model in the large-$q$ limit. The results demonstrate that the leading Lyapunov exponent in the chaotic phase  tends toward zero at large times.
In Sec.~\ref{sec:mbl_phase}, we discuss the Lyapunov exponents for models of non-interacting systems, systems with small coupling and systems with local integrals of motion. In this way we argue that the leading Lyapunov exponent remains finite at large $t$ in the MBL phase. 
In Sec.~\ref{sec:numerics}, we present numerical results for the two quantum circuit models. Within the limitations imposed by the maximum computationally accessible values of $t$, results are consistent with distinct types of behaviour in each phase as described above.
Finally, we conclude and discuss the outlook in Sec.~\ref{sec:dis}.

\section{Spectral Lyapunov exponents and generating function}\label{sec:def}
Consider an ensemble of disordered systems, each associated with a Floquet operator $W$ which we assume spatially inhomogeneous due to the presence of local disorder.  
Let  $\{ \theta_m \}$ be the eigenphases of $W$. We introduce the higher point spectral form factor (hpSFF)\cite{complexity2017,liu2018}  as 
\begin{equation}\label{eq:hpsff}
 \langle \mathcal{K}_L(t)^\alpha \rangle :=  
 \left\langle 
 \bigg[
 \sum_{m,n}  e^{ \imath (\theta_m - \theta_n)t } 
 \bigg]^\alpha
  \right\rangle = \langle |\Tr[W(t)]|^{2\alpha} \rangle\; ,
\end{equation}
where $\langle \cdot \rangle$ is the ensemble average, and the subscript denotes the system size $L$ with periodic boundary conditions.
For $\alpha = 1$, we have the standard SFF,  $K_L(t) = \langle \mathcal{K}_L(t)\rangle$. 
To study fluctuations of the hpSFF (which are exponentially large in $L$) in the thermodynamic limit, we introduce the scaled cumulant generating  function
\begin{equation}
\label{eq:Fdef}
 F_t(\alpha) \equiv \lim_{L \to \infty}\frac{1}{L}\log \langle \mathcal{K}_L(t)^\alpha \rangle \;.
\end{equation}
\begin{figure}[ht]
\includegraphics[width=0.45\textwidth]{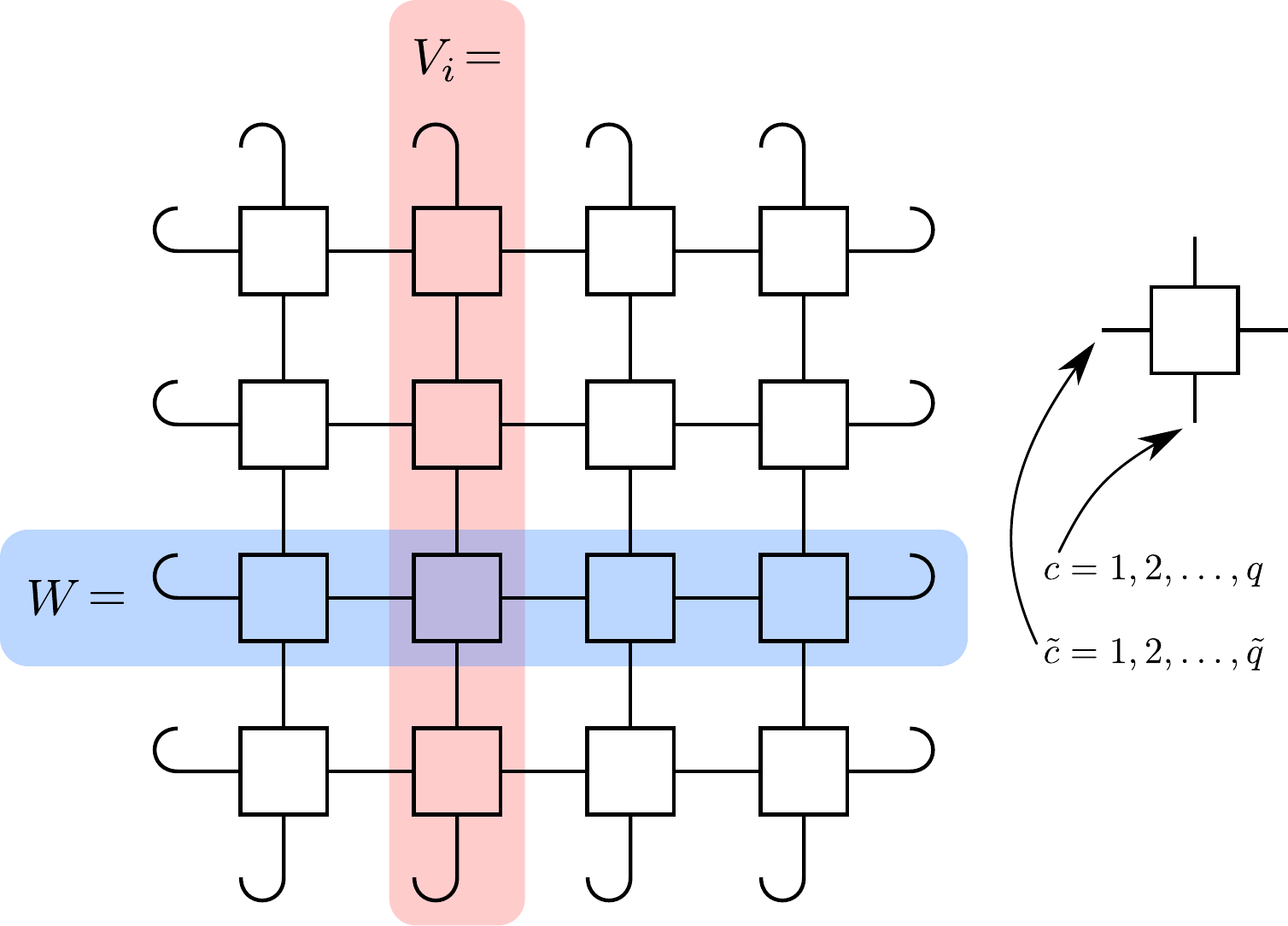}
\caption{Left: A diagrammatic representation of Eq.~\eqref{eq:dual}. The operator $W$ is represented as a matrix product operator (MPO). The curly lines on the top and bottom boundaries represent a trace of $W$, and  ones on the left and right boundaries represent periodic boundary condition.
Right: Each MPO acts vertically on a physical space with dimension $q$, and horizontally on an auxiliary space with dimension $\tilde{q}$. 
\label{Fig:diagrdual}}
\end{figure}
As we will see below, the function $F_t(\alpha)$ captures the large-$L$ scaling of all cumulants of the SFF. Knowledge of it gives access to the large-deviation distribution of $K(t)$.  
By definition, $F_t(\alpha)$ is a convex function. 

The behavior of $F_t(\alpha)$ can be analysed by considering a dual picture \cite{GutkinOsipov,Akila}, using a 90-degree rotation which exchanges space and time. To be more concrete without losing generality, we can represent $W$ as a matrix-product operator, where the vertical bonds have the physical dimension $q$ and the \textit{auxiliary} horizontal ones have dimension $\tilde q$. Then we can rewrite its trace in the dual picture as
\begin{equation}
\label{eq:dual}
\Tr_{\mathcal{H}}[W(t)] = \Tr_{\mathcal{\tilde H}}[V_L V_{L-1} \ldots V_1] = 
\Tr_{\mathcal{\tilde H}} [V^{(L)}] \;,
\end{equation}
where the operators $V_i$ are $\tilde q^t \times \tilde q^t$ matrices defined implicitly by the diagram in Fig.~\ref{Fig:diagrdual}. To avoid confusion, we have written explicitly the Hilbert space where the trace is taken as a subscript and we set $V^{(L)} \equiv V_L \ldots V_1$. 
Since the operator $W$ is inhomogeneous in space, the matrices $V_i$ are different one from the other and randomly distributed due to the presence of local disorder. 

At this stage, one can proceed in two ways. One possibility is to perform the average over the disorder by considering $2\alpha$ layers $W(t) \otimes \ldots \otimes W(t) \otimes W^{\dag}(t) \otimes \ldots W^{\dag}(t)$. By using Eq.~\eqref{eq:dual}, this amounts in practice to computing the disorder average of $2\alpha$ replicas of the single-slice transfer matrix $V_i$ for integer $\alpha$. The resulting transfer matrix leads directly to $\langle\mathcal{K}(t)^\alpha\rangle$. 
Additionally, after averaging, the resulting transfer matrix is invariant under spatial translations and so it is sufficient to study a single slice, and its leading eigenvalues and associated eigenvectors. 
This approach was employed recently in several studies~\citehere{cdc2, bertini2018exact, Guhr2019, flack2020statistics, garratt2020manybody, garratt2020MBL, lerose2020influence, sonner2020characterizing}.
%
%
We will use this method to compute analytically $F_t(\alpha)$ in the limit of large local Hilbert space dimension.

Another possibility is to consider the transfer matrix for a single layer $W(t)$. This has the advantage for numerical calculations that its size (${\tilde q}^t \times {\tilde q}^t$) is smaller and independent of $\alpha$. However, since there is no sense in averaging $W(t)$, we have to study this transfer matrix for individual samples. That means at large $L$, $V^{(L)}$ is the product of many random matrices. The natural quantities that characterise this product are the Lyapunov exponents. 
More precisely, in order to define them, we note that the trace in Eq.~\eqref{eq:dual} enforces periodic boundary conditions and homogeneity in time ensures that the matrices $V_i$ are invariant under translations in the time directions. There is therefore a momentum quantum number associated with the time direction. 
The spectral decomposition of $V^{(L)}$ can thus be organised into the different momentum sectors $k = 2\pi j/t$, with $j = 0,\ldots t-1$, in the form
\begin{equation}
\label{eq:lyap_finiteL}
 V^{(L)} = \sum_k \sum_{a} \ket{\ell_a (k)} e^{\lambda_{a}^{(L)}(k) L/2 + \imath \phi_{a}^{(L)}(k)} \bra{r_a(k) }
 \;, 
\end{equation}
where $\lambda_0^{(L)}(k) \geq \lambda_1^{(L)}(k) \geq \ldots$ are  growth rates which have sample-to-sample fluctuations for finite $L$ but converge with probability one to the \textit{spectral Lyapunov exponents} with momentum $k$. We refer below to these growth rates as finite-size \textit{spectral Lyapunov exponents}.  The $\phi_{a}^{(L)}(k)$'s are the corresponding phases, while $\ket{\ell_a}$ and $\ket{r_a}$ are respectively the left and right eigenvectors, which are biorthogonal and normalized such that $\braket{ r_a| \ell_b}  =\delta_{ab}$.
We find that the largest Lyapunov exponent always lies in the zero-momentum sector, so for convenience we denote
\begin{equation} 
    \lmax^{(L)} \equiv \lambda_{0}^{(L)} (k=0) \; .
\end{equation}
Furthermore, for any finite $t$, there is always a gap $\Delta \lambda^{(L)}$ between $\lambda_>^{(L)}$ and the other Lyapunovs, so that at large $L$
\begin{multline}
\label{eq:Klyap}
 \mathcal{K}_L(t) = 
 \sum_{k,k'}\sum_{a,a'} e^{ (\lambda_{a}^{(L)} (k)   
 + \lambda_{a'}^{(L)}  (k)) L/2} 
 e^{\imath (\phi_{a}^{(L)} (k) - \phi_{a'}^{(L)} (k'))} \\ 
\sim
e^{\lmax^{(L)} L} + O(e^{-L \Delta\lambda})
\;.
\end{multline}

From Eq.~\eqref{eq:Fdef} it follows that 
\begin{equation}
 F_t(\alpha) = \lim_{L \to \infty} \frac 1 L \log \langle e^{\alpha \lmax^{(L)} L}\rangle \;,
\end{equation}
and that derivatives of $F_t(\alpha)$ at $\alpha = 0$ provide cumulants of the largest finite-size Lyapunov exponent: 
\begin{equation} \label{eq:F_cumulants}
\left. \frac{d^n}{d{\alpha}^n} F_t(\alpha) \right|_{\alpha = 0} = \lim_{L \to \infty} L^{n-1} \langle [\lmax^{(L)}]^n \rangle_{\mathrm{c}}  \;.
\end{equation}
In particular, when $L \to \infty$, we extract the average and variance 
\begin{align}
\label{eq:lambda_inf}
F'_t(\alpha = 0) &= \lim_{L\to\infty}\langle \lmax^{(L)} \rangle \equiv \lmax  \;, \\
F''_t(\alpha = 0) &= \lim_{L\to\infty} L  \mbox{var}(\lmax^{(L)}) \;.
\end{align}
Therefore, provided $F_t''(\alpha = 0)$ is not divergent, in the limit $L \to \infty$, the distribution of $\lmax$ is concentrated on its mean $\lmax$ almost surely and the function $F_t(\alpha)$ encodes its large deviations. In contrast with $\mathcal{K}_L(t)$, the Lyapunov exponents are thus self-averaging. In general we will denote
\begin{equation} \label{eq:lyap_k}
    \lambda_a(k) = \lim_{L\to\infty} \langle \lambda_a^{(L)}(k) \rangle \;,
\end{equation}
which defines the spectral Lyapunov exponents.
In the following, we will study the $k$-dependence of the leading Lyapunov exponents $\lambda_0 (k)$ and the fluctuations of $\lmax^{(L)}$ as encoded by the generating function $F(\alpha)$ for chaotic and MBL systems.

\section{Models \label{sec:models}}
For our analytical and numerical analysis we will consider two main models: the random phase model (RPM)~\cite{cdc2} and the kicked Ising model (KIM)~\cite{bertini2018exact,Guhr2019}.
Below, we summarise their definitions and main features. 
\subsection{Random Phase Model (RPM) \label{subsec:RPM}}
The RPM consists of $q$-state `spins' arranged with nearest-neighbour coupling on a one-dimensional lattice. 
We use site labels $n=1 \ldots L$ and orbital labels $a_n = 1 \ldots q$ on the $n$-th site. The $q^L\times q^L$ Floquet operator $W=W_2\cdot W_1$ is a product of two factors. 
\begin{equation}
\label{eq:W1}
W_1 = U_1 \otimes U_2 \otimes  \ldots U_L
\end{equation}
generates rotations at each site $n$, with $q\times q$ unitary matrices $U_n$ chosen randomly and independently from the circular unitary ensemble (CUE). $W_2$ couples neighbouring sites and is diagonal in the basis of site orbitals. 
The phase of the diagonal elements is a sum of terms depending on the quantum states of adjacent sites, so that
\begin{equation}
\label{eq:W2}
[W_2]_{a_1, \ldots a_L;a_1, \ldots a_L} = 
\exp\left(\imath \sum_n \varphi_{a_n,a_{n+1}}^{(n)} \right)\,.
\end{equation}
We take each $\varphi^{(n)}_{a_n,a_{n+1}}$ to be an independent Gaussian random variable with 
mean zero and standard deviation $\epsilon$, which effectively controls the coupling between neighbouring spins.

For fixed $q$, the model exhibits a many-body localization transition as a function of $\epsilon$~\cite{cdc2, mac2019quantum}, with a critical value $\epsilon_c$ separating an MBL ($\epsilon < \epsilon_c$) from a chaotic phase ($\epsilon > \epsilon_c$). We will employ this model for exact analytic calculations within the chaotic phase, in the limit $q \to \infty$. Note that accessing the MBL phase in this limit is problematic as $\epsilon_c \to 0$ when $q\to \infty$. We will therefore complement the analysis with numerical studies at $q=3$, for which the model has  $\epsilon_c\approx 0.25$~\cite{cdc2}. 

\subsection{Kicked Ising Model (KIM)}\label{subsec:KIM}
The kicked Ising Model (KIM) is a Floquet Ising spin-$1/2$ chain defined by the time evolution operator $W= W_2 \cdot W_1$ with
\begin{subequations}
	\label{kickedspin}
	\begin{align}
	& W_1 = e^{\imath \sum_{i} h_i \sigma_i^z} e^{\imath \sum_{i} b \sigma_i^x} \label{sigmazdef} \;,
	\\
	& W_2 = e^{\imath \sum_{i} J \sigma_i^z \sigma_{i+1}^z} \;,
	\end{align}
\end{subequations}
with $h_j, J$ and $b$ real parameters. 
Similarly to the RPM, this model has a many-body localization transition at a critical coupling strength $J_c= 0.23$~\citehere{Guhr2019}, so that it exhibits a MBL phase for $J < J_c$ and a chaotic phase for $J_c <J \leq \pi/4$.
This model has recently received a lot of attention, because of the existence of a ``self-dual point'' in the parameter space: $|J|= |b|=\pi/4$ and arbitrary local longitudinal fields $h_j$. For these special values of the parameters,  not only the evolution operator $W$, 
but also its duals $V_j$ (see Eq.~\eqref{eq:dual} and Appendix \ref{app:dual} for the detailed definition) acting in the space direction, can be chosen to be unitary and with the same form of Eq.~\eqref{kickedspin}. In this case, at all times $t$, not only the average SFF~\cite{bertini2018exact}, but also its higher moments are in perfect agreement with the prediction of an appropriate random matrix ensemble which takes care of all the symmetries~\cite{flack2020statistics}. 
Indeed,
unlike the RPM, this model is time-reversal invariant, and consequently, the behaviour of SFF for $ t_{\mathrm{Hei}} > t \gtrsim \tth$ is expected to follow the circular orthogonal ensemble (COE), which is given in the limit of large random matrices for $t\ll t_{\mathrm{Hei}}$ by 
\begin{equation}
\label{eq:SFFCOE}
K_{\mathrm{COE}} \sim 2t 
\;.
\end{equation}
 At the self-dual point, additional discrete symmetries have been identified for the dynamics induced by Eq.~\eqref{kickedspin}, but they become irrelevant at large $t$~\cite{flack2020statistics, bertini2018exact, Guhr2019}.   

Although solvable, the behavior at the self-dual point is not generic~\cite{kos2020correlations}, as it implies for instance that $\tth$ does not diverge with the system size but remains $O(1)$.
Here, we will mainly use this model for numerical analysis without restricting to the self-dual point, taking advantage of its particularly small finite-time corrections near the self-dual point.

\section{The chaotic phase}\label{sec:eth_phase}
\subsection{General behavior}
We start by focusing on systems belonging to the CUE symmetry class and on the case $F_t(\alpha = 1)$, which is simply related to the usual average of the spectral form factor $\langle \mathcal{K}(t)\rangle$. We make use of the defining property Eq.~\eqref{eq:Fdef} to estimate the behavior of $F_t(\alpha=1)$ at large $t$ in the chaotic phase. As observed in \citehere{cdc2, friedman2019, moudgalya2020spectral,garratt2020manybody}, for systems in the CUE symmetry class, the SFF  approaches the random matrix prediction 
\begin{equation}
\label{eq:Fkt}
    \langle \mathcal{K}_L(t)\rangle \sim K_{\rm CUE}(t) = t \;, \quad t \gtrsim \tth(L)  \;.
\end{equation}
The specific details controlling the behavior $\tth(L)$ are not yet fully understood, but in different set-ups\cite{cdc2, friedman2019, moudgalya2020spectral} one expects $\tth \propto L^{\nu}$, with $\nu>0$\footnote{A logarithmic scaling $t\propto \log L$ has been observed in the $q\to \infty$ of the RPM, but power-law is expected at finite $q$, see Sec.~\ref{sec:RPMqinf} and \citehere{cdc2}.}. As a consequence, as already stated, the Thouless time $\tth(L) \to \infty$ when $L \to \infty$.
Although $F_t(\alpha)$ is formally defined only in the limit $L\to\infty$, we expect it to capture well the finite-$L$ behavior of $\mathcal{K}_L(t)$ when $t \sim \tth(L)$. From Eq.~\eqref{eq:Fdef},
we can write
\begin{equation}
K_L(t) = \exp[L (F_t(\alpha) + o(1))]
\end{equation}
and this suggests that, in order for the exponential growth in $L$ of $K_L(t)$ to be suppressed, we must have $F_{\tth}(\alpha = 1) \lesssim 1/L$. We thus deduce the scaling
\begin{equation}
    F_{t}(\alpha = 1) \lesssim t^{-1/\nu} \;.
\end{equation}
This argument can be extended to other values of $\alpha$ and we reach the conclusion that the chaotic phases must be characterized by
\begin{equation}
\label{eq:Ftchaos}
    \lim_{t \to \infty} F_t(\alpha) = 0 \;, \qquad \forall \alpha\geq 0 \;.
\end{equation}
In the next subsection, we will quantitatively justify this statement by computing explicitly $F_t(\alpha)$ for the RPM in the limit $q\to\infty$.

Additionally, we see that not only the leading Lyapunov in the zero-momentum sector $\lmax$, but $t$ Lyapunov exponents have to vanish in the large-$t$ limit in order to reproduce the linear growth in time in \eqref{eq:Fkt}. 
The most natural assumption is that the $t$ vanishing Lyapunov exponents correspond to the different $\lambda_0(k)$ in the $t$ momentum sectors. Similarly, for systems belonging to the COE symmetry class, in order to fulfill Eq.~\eqref{eq:SFFCOE}, we expect \emph{two} vanishing Lyapunov exponents for $t \to \infty$ in each momentum sector.
We support these conjectures with numerical simulations in Sec.~\ref{sec:numerics}

\subsection{$F_t(\alpha)$ for RPM at $q\to\infty$
\label{sec:RPMqinf}}

As a solvable model of the chaotic phase in a spatially extended many-body quantum system we consider the RPM \cite{cdc2}, 
and compute analytically $F_t(\alpha)$ and $ \lambda_> $ in the large-$q$ limit.
We first of all consider integer values of $\alpha = \thetaint$. We map the computation of $\langle \mathcal{K}(t)^\thetaint \rangle$ to the partition function of a one-dimensional statistical mechanical problem with nearest-neighbour interactions. As explained above, we introduce a transfer matrix in the space direction which allows the exact computation of $\langle K(t)^\thetaint \rangle$ in the limit of large $q$: 
the value of $F_t(\thetaint)$ corresponds to the leading eigenvalue of the transfer matrix for the statistical mechanics problem, in a way that generalises the approach described in \citehere{cdc2}. Lastly, we analytically continue $F_t(\alpha)$ to non-integer $\alpha$ and obtain $ \lambda_> $.
\begin{figure}[htb]
	\includegraphics[width=0.42\textwidth]{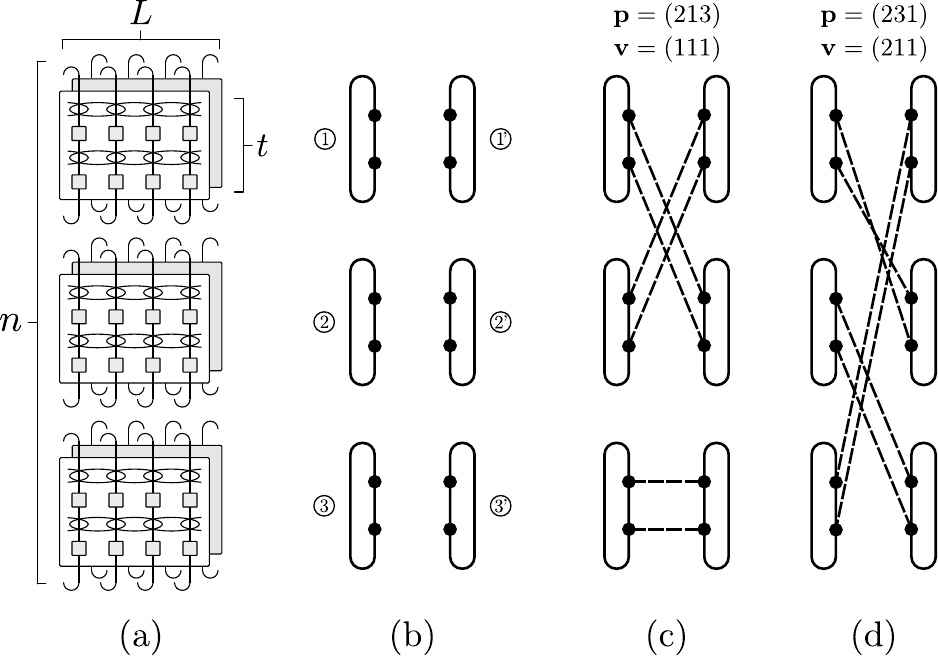}
	\caption{Construction of the Hilbert space associated with the transfer matrix for a statistical mechanics problem. (a) Diagrammatic representation of $\langle \mathcal{K}(t)^\thetaint \rangle$. Space and time are represented by the horizontal and vertical directions. The boxes and ellipses represent the Haar-random 1-gates and the diagonal 2-gates respectively. The white and grey sheets represent $W(t)$ and $W^\dagger (t)$ respectively. The curly lines on top and bottom represent traces. (b) Diagrammatic representation of a single site, where the 2-gates are omitted. (c) and (d): Two examples of single site configurations after the ensemble average over $W_1$ in the large-$q$ limit.}
	\label{fig:rp_w1}
\end{figure}

To derive the transfer matrix  for $\langle \mathcal{K}(t)^\thetaint \rangle$, we construct the associated Hilbert space by performing the Haar-average over $W_1$ for each site independently, as illustrated in Fig.~\ref{fig:rp_w1}. 
This independent averaging over $W_1$ is legitimate because the 1-gates are drawn independently across different sites, and because $W_2$ consists of \textit{diagonal} 2-gates only. 
Using the procedure explained in \citehere{cdc1} and \citehere{cdc2}, we find a total of $n! \, t^n$ diagrams at each site in the limit of large-$q$. To each diagram we associate a state in the Hilbert space, labelled by a vector $\mathbf{v}$ in $\mathbb{Z}^\thetaint_t$ and by $\mathbf{p} = (\sigma(1), \sigma(2) \dots \sigma(\thetaint))$ where $\sigma$ belongs to the permutation group $S_\thetaint$ of $\thetaint$ elements. 
Fig.~\ref{fig:rp_w1}a is the diagrammatic representation of $\langle \mathcal{K}(t)^\alpha \rangle$.  Fig.~\ref{fig:rp_w1}b is the diagrammatic representation of a given site where each Haar-random 1-gate is represented by a single dot. \footnote{Only a single dot is used since the diagrams are shown to be ``Gaussian'' in the large-$q$ limit \cite{cdc1}.}
Upon averaging, the $j$-th loop (out of $\thetaint$ loops) on the left is paired with the $p_j$-th loop on the right in Fig.~\ref{fig:rp_w1}b. Furthermore, the  pairing of $j$-th loop will have 1 out of $t$ possible configurations, labelled by $v_j$. Fig.~\ref{fig:rp_w1}c and d are two examples.

The average over $W_2$ in the large-$q$ limit gives the matrix elements of the transfer matrix $T$
\begin{equation}\label{eq:rp_tmat}
\bra{\mathbf{p}, \mathbf{v}}T (t,\thetaint) \ket{\mathbf{p}', \mathbf{v}'} 
=
\exp(-\epsilon \, \thetaint \, t)
\exp\Bigg[ \epsilon t \sum_{j =1}^\thetaint \delta_{p_j, p'_j} \delta_{v_j, v'_j}
\Bigg]  \;,
\end{equation}
which is constructed by counting the unmatched configurations and pairings between configuration $(\mathbf{p}, \mathbf{v})$ and $(\mathbf{p}', \mathbf{v}')$, since each unmatched configuration gives a factor of $\exp(-\epsilon \thetaint t)$. As an example, the matrix element between the states in Fig.~\ref{fig:rp_w1}c and d is $\exp(- 6 \epsilon)$, since $\thetaint = 3$, $t=2$ and none of the pairings or configurations match. In summary, we have shown that the evaluation of $\langle K(t)^\thetaint \rangle$ can be mapped to a one-dimensional statistical mechanical model where each site has $n!t^n$ states and where the interaction is defined by Eq.~\eqref{eq:rp_tmat}.

\begin{figure}[t!]
	\includegraphics[width=0.95\columnwidth]{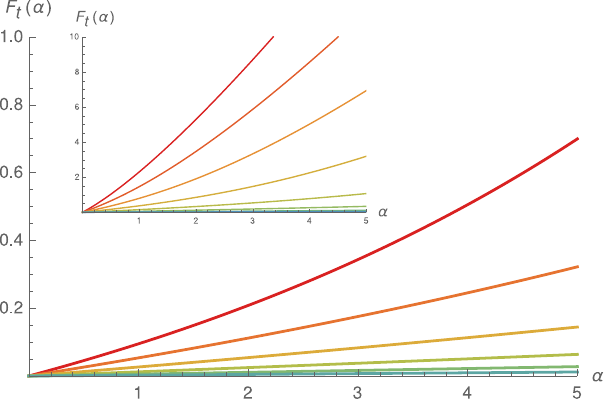}
	\caption{Main panel: Large-$q$ analytical results for $F_t(\alpha)$ vs $\alpha$ for different $t$ at fixed $\epsilon=1$ for the RPM. 
	The rainbow colours correspond to different values of $t$, from $t=3$ in red to $t=8$ in blue in steps of 1. Solutions for $t>8$ are very small on the scale shown.  
	Inset: Large-$q$ results for $F_t(\alpha)$ vs $\alpha$ at $t=10$ for the RPM.  The rainbow colours correspond to different values of $\epsilon$, from $\epsilon=0$ in red to $\epsilon=0.8$ in steps of $0.1$. Results for $\epsilon \gtrsim 0.8$ are very small on the scale shown.  
	} \label{fig:rpm_f_vs_alpha_diff_t_diff_eps}
\end{figure}

\begin{figure}[t!]
	\includegraphics[width=0.95\columnwidth]{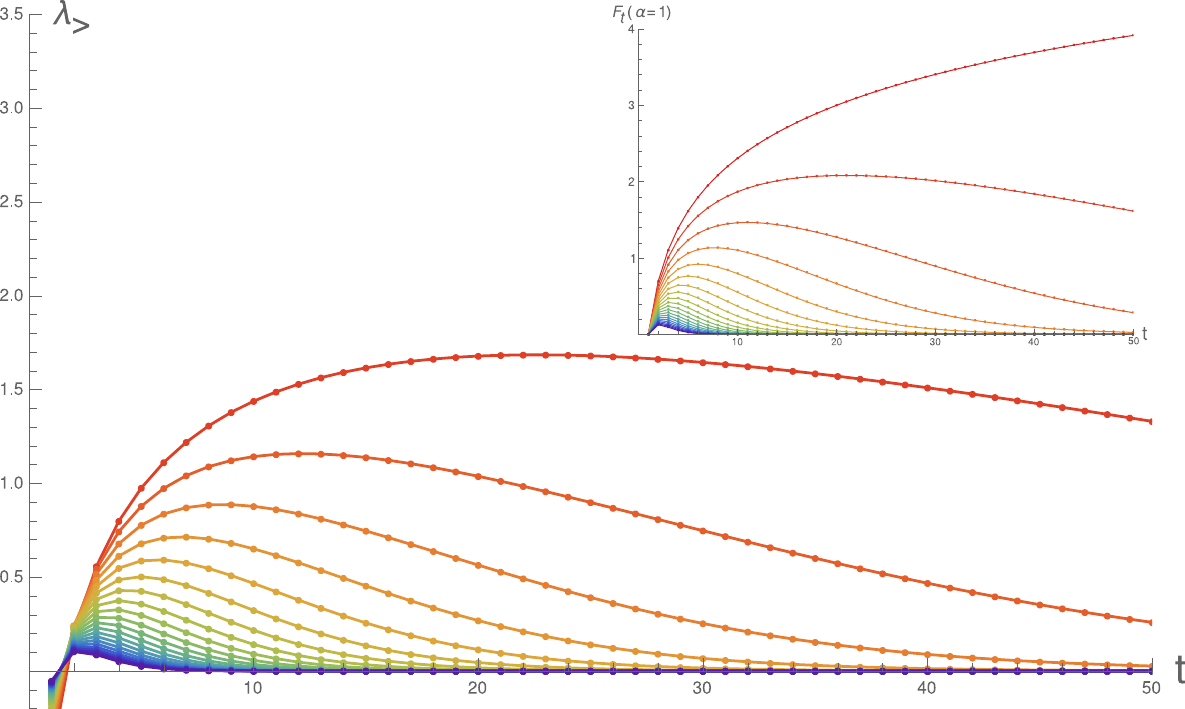}
	\caption{
	Main panel:
	Large-$q$ analytical results for $ \lambda_> $ vs $t$ for the RPM. The rainbow colours correspond to different values of $\epsilon$, from $\epsilon=0.05$ in red to $\epsilon=1$ in blue in steps of $0.05$.   
Inset: Large-$q$ analytical results for $F_t (\alpha=1)$ vs $t$ for the RPM.   The rainbow colours correspond to different values of $\epsilon$, from $\epsilon=0$ in red to $\epsilon=1$ in blue in steps of $0.05$.
%
%
%
 \label{fig:rpm_lambda_vs_t}
}
\end{figure}

In Appendix~\ref{app:F_eth}, we compute the leading eigenvalue $E_t(\alpha)$ of $T$ and analytically continue the result from integer $n$ to arbitrary $\alpha \geq 0$ to obtain  
\begin{equation}
E_t(\alpha) = 
e^{\frac{1-x}{t x}} (t x)^{\alpha } \Gamma \Bigl(\alpha +1,\frac{1-x}{t x}\Bigr) \;,
\end{equation}
where $\Gamma(a,b)$ denotes the incomplete Gamma function and we parameterize $x = e^{- t \epsilon }$. As a consistency check, at $\epsilon=0$, $E_t(\alpha) = \alpha! t^\alpha $ as expected since all the entries of the transfer matrix are unity. 
In general, we have the relation
\begin{equation}
\lim_{q\to \infty} F_t(\alpha) = \log E_t(\alpha) \;.
\end{equation}
At large times  ($x \ll 1$), we obtain the expansion
\begin{equation}\label{eq:F_small_x}
F_t(\alpha) = \alpha  (t-1) x  + \frac{1}{2} x^2 ((\alpha^2 -2\alpha)  t^2+2 \alpha  t - \alpha) + O(x^3) \;.
\end{equation}
Using the replica trick, the Lyapunov exponent can be computed as 
\begin{equation} \label{eq:lambda_rpm}
\lmax = 
e^{\frac{1-x}{t x}} \Gamma\left(0,\frac{1-x}{t x} \right)
+\log \left(1-x \right) \;.
\end{equation}
These analytical solutions are plotted in Figs.~\ref{fig:rpm_f_vs_alpha_diff_t_diff_eps} and \ref{fig:rpm_lambda_vs_t}. Fig.~\ref{fig:rpm_f_vs_alpha_diff_t_diff_eps} shows that $F_t(\alpha)$ becomes flat as $t$ (main panel) and $\epsilon$ (inset) increase, which implies $ \lambda_> $ tends to zero for increasing $\epsilon$ and $t$. This behaviour of $ \lambda_> $ in time is shown more explicitly in Fig.~\ref{fig:rpm_lambda_vs_t}. 
Note that for small $t$ (in particular $t=1$) $\lambda_>$ is negative. We will see that this short-time feature also appears in the finite-$q$ numerics.
Fig.~\ref{fig:rpm_lambda_vs_t} inset shows the analytic result for the log of the averaged SFF.

We use Eq.~\eqref{eq:F_small_x} to define a generalized Thouless time $\tthg$ associated with the hpSFF as the time after which hpSFF behaviour (of a quantum many-body system) coincides with the RMT result. 
For the CUE in the large-$q$ limit, the hpSFF is exactly $\alpha! t^\alpha$ due to the same diagrammatic approach explained in Fig.~\ref{fig:rp_w1}. The transfer matrix \eqref{eq:rp_tmat} becomes the identity matrix in the limit of large-$t$, and its trace gives the hpSFF CUE result as expected. To compute the $\tthg$, we demand the $L$-th power of the leading Lyapunov exponent to be $O(1)$, i.e. $ F_t(\alpha) \sim 1/L$. Using Eq.~\eqref{eq:F_small_x}, we see that $\tthg = O(\log L)$ independent of $\alpha$. This result generalises the logarithmic scaling obtained in \citehere{cdc2} at $\alpha = 1$.

\section{The MBL phase}\label{sec:mbl_phase}
In this section, we discuss the general features of $F_t(\alpha)$ and of the Lyapunov spectrum $\lambda_0(k)$ in the MBL phase. 
In order to obtain some intuition, we first treat the case of uncoupled sites by analyzing the RPM at $\epsilon = 0$. Then we consider systems with small coupling using a perturbative analysis applicable to both the RPM and KIM. Lastly, we analyze the leading Lyapunov exponent for an effective model of MBL in terms of local integrals of motion (LIOM).   
\subsection{Uncoupled sites}
We use the RPM at $\epsilon = 0$ as a toy model for the MBL phase. In this case $W_2$ 
is simply the identity and the model reduces to $L$ non-interacting spins, each independently evolving with a random CUE matrix. From Eq.~\eqref{eq:hpsff}, we obtain for all moments
\begin{equation}
\label{eq:Knonint}
    \langle \mathcal{K}_L(t)^\alpha \rangle 
  = \langle |\Tr[U(t)]|^{2\alpha} \rangle_{\rm CUE}^L\; ,
\end{equation}
where the average is performed within the CUE from which $ U$ is drawn. Given the trivial dependence of \eqref{eq:Knonint} on the system size $L$, we see from \eqref{eq:Klyap}  that  except for $\lambda_0(k=0)$, all the other Lyapunov exponents (thus including all $\lambda_0(k \neq 0$)) are degenerate with the value $-\infty$: this is a general feature of models with uncoupled sites. 
From \eqref{eq:Fdef}, we obtain an expression for  $F_t(\alpha)$ in terms of average of a single CUE matrix. In particular, for $q = 2$, we obtain the explicit formula
\begin{equation}
\label{eq:singleq2}
    \lim_{t \to \infty} F_t(\alpha) = \ln \left[ \frac{4^\alpha \Gamma(\alpha + 1/2)}{\Gamma(\alpha +1 )}\right] \;, \quad q = 2 \;.
\end{equation}
Note that the large time limit washes away many microscopic details and this expression holds more generally for non-interacting spins $1/2$ with an arbitrary distribution of random fields, thus including the KIM 
at $J = 0$, as well as disordered free fermions in one dimension, i.e. the Anderson model~\cite{RevModPhys.80.1355}.
For $q > 2$, one cannot get an analytic expression; nevertheless at large $q$ but $t \gg q$, one can use that $\Tr[U(t)]$ behaves as a gaussian-distributed complex random variable with zero average and variance $q$, leading to
\begin{equation}
\label{eq:singlelargeq}
    F_t(\alpha) \sim \ln [q^\alpha \Gamma(\alpha + 1)] \;, \qquad t \gg q \gg 1
\end{equation}
Note that in realistic models, the limit of large time is reached quite quickly, whenever $t$ is larger than the single-spin Heisenberg time, i.e. $t \gg q = O(1)$.

By contrasting \eqref{eq:singleq2} and \eqref{eq:singlelargeq} with \eqref{eq:Ftchaos}, we observe a first indication of the different behaviour in a non-ergodic phase: $F_t(\alpha)$ converges to a non-zero function at large $t$. In the next sections we will see that this feature also characterises the MBL phase.

While accessing numerically the whole function $F_t(\alpha)$ can be problematic, we will show in Sec.~\ref{sec:numerics} that the neighbourhood of $\alpha = 0$ can be studied efficiently. Indeed, with the exception of special cases (e.g. for non-interacting spin $1/2$, Eq.~\eqref{eq:singleq2} leads to $F'_t(0) \stackrel{t\to \infty}{\longrightarrow} 0$), the behaviour $F'_t(0) \neq 0$ at large times provides a sufficient indication of a non-ergodic phase.

\subsection{Perturbative analysis at small coupling}
The dual transfer matrix provides an interesting framework in which to perform a perturbative expansion at small coupling between sites. The technique can be applied to both the KIM and the RPM, but we focus on the first. In Appendix  \ref{app:dual} and \ref{app:pert}, we show that the transfer matrix $V_i$ corresponding to the two-layer structure introduced in Eqs.~\eqref{kickedspin} can be written as $\VV_i = {\VV}^{(1)}_i{\VV}^{(2)}_i$
where 
\begin{subequations}
\label{eq:dualspinshort}
\begin{align}
&\VV_{1} \equiv 
\prod_{\mu=1}^t (e^{\imath J}  {\mathbf{1}}_{\mu} + e^{-\imath J} {\sigma}_\mu^x) \;,
\\
&\VV_{2,i} \equiv \left[\frac{\imath}{2} \sin(2 b)\right]^{t/2} e^{\imath h_j \sum_{\mu=1}^t \sigma_\mu^z  + \ffun(b) \sigma^z_\mu 
\sigma^z_{\mu+1}} \; ,
\end{align}
\end{subequations}
and $\ffun(a) = \operatorname{arctanh}(e^{-2 \imath a})$. Note the resemblance with Eqs.~\eqref{kickedspin} whose unitary form is recovered at the self-dual unitary point $|b| =|J| = \pi/4$~\citehere{bertini2018exact}. Here, we focus on $b = \pi/4$ and small $J$. The operator ${\VV}_1$ is easily diagonalised, 
and at small $J$ the leading eigenstate is $\ket{\bf 0} = \ket{+\ldots +}$, with $\sigma^x \ket{\pm} = \pm \ket{\pm}$. Every spin flip $\sigma^z_j \ket{\bf 0}$ is suppressed by a power of $J$.
At the leading order in $J$, we thus truncate the Hilbert space of the trace in \eqref{eq:dual} to states only involving up to one spin flip $\sigma^z_\mu \ket{\bf 0}$, where we use Greek letters $\mu = 0,\ldots, t-1$ to parameterise the position in time in the dual Hilbert space. Additionally, we employ the translational invariance in the time direction so that, within this truncation, we have a single \textit{magnon} in each momentum sector $k$
\begin{equation}
\label{eq:magnon}
    \ket{k} = \frac{1}{\sqrt{t}} \sum_{\mu = 0}^{t-1} e^{\imath k \mu} \sigma^z_j \ket{\bf 0} \;, \quad k = \frac{2 \pi n}{t} \;.
\end{equation}
We can thus obtain an expression for $\lambda_0(k)$ for every $k \neq 0$, which takes the form (see Appendix \ref{app:pert} for the full derivation)
\begin{multline}
\label{eq:pertfinal}
    \lambda_0(k) = \overline{\ln |\braket{k |  V_2 | k}|^2} + \ldots \sim  2 \ln |J| +  \\ 
    \int dh P(h) \ln \left[\frac{\cos(h)^2}{(2 - \cos(h)^2) (\cos k + \sin(h)^2)^2 }\right] + O(J^2)  \;, 
\end{multline}
where $P(h)$ is the probability distribution of the random fields $h_i$. A comparison between Eq.~\eqref{eq:pertfinal} and numerically exact results is shown in Fig.~\ref{fig:lam_vs_k_kim}. Note that at first order in $J$, the time variable does not appear explicitly in Eq.~\eqref{eq:pertfinal}. We can thus take the $t\to\infty$ limit, where $k$ becomes a continuous variable $k\in[-\pi,\pi]$. We leave for further investigation the study of the convergence of higher order corrections, but quite interestingly Eq.~\eqref{eq:pertfinal} provides an explicit result in the limits of both large times and large system sizes.

The case $k = 0$ needs a different treatment because even at the leading order, $O(J)$, the zero-momentum sector is two-dimensional, containing $\ket{\bf 0}$ and the zero-momentum magnon $\ket{k = 0}$ in Eq.~\eqref{eq:magnon}. This fact is at the origin of the discontinuity observed in the spectrum at $k = 0$ (see Fig.~\ref{fig:lam_vs_k_kim}).
The resulting Lyapunov exponents $\lambda_0(0)$ and $\lambda_1(0)$ cannot be written analytically but can easily be computed numerically (see Appendix \ref{app:pert}).

\subsection{Local Integrals of Motion}
To describe the general behavior of the (fully) MBL phase, we consider an effective model based on the hypothesis that the MBL phase is characterised by an extensive number of  LIOM with exponentially decaying interactions \cite{serbyn13mbl,huse14mbl},
\begin{equation}
H = \sum_{i} J^{\mathrm{(1)}}_i \tau^z_i 
+ \sum_{i < j} J^{\mathrm{(2)}}_{i,j}  \tau^z_i \tau^z_{j} 
+ \sum_{i < j <k } J^{\mathrm{(3)}}_{i,j,k}  \tau^z_i \tau^z_{j}  \tau^z_{k} 
+ \dots
\;,
\label{eq:liom}
\end{equation} 
where the operators $\tau^\alpha_i$ with $\alpha = x,y,z$ form a spin $1/2$ representation for each $i$ but 
have an exponentially-decaying support in real space around the physical site $i$, i.e. $||[\tau^z_i, \sigma^z_j]|| = O(e^{- |i - j|/\xi})$, with $\xi$ the localization length, and $||\cdot||$ the operator norm.
The $\tau^z_i$ provide an extensive set of integrals of motion that do not relax. Relaxation for real spins $\sigma_i^z$ operators is thus induced by the accumulating random phases between different components of the system. This dephasing dynamics in MBL is the origin of logarithmic growth of entanglement\cite{bardarson2012,serbyn2013} and power-law relaxation of local observables\cite{serbyn2014a,serbyn2014b}.

For simplicity, we focus on the two-body model where $J^{(n)}_{i,j,\dots}=0$ for $n\geq 3$, and where $J^{(2)}_{i,i+r}$ are independently and Gaussianly distributed for each $i$ and $r$, i.e.
\begin{align}\label{eq:j1j2}
\langle  (J_{i}^{(1)})^2\rangle = J_1^2 \, , 
\quad
\langle  (J_{i,i+r}^{(2)})^2\rangle = J_{2,r}^2 
\;.
\end{align}  
Furthermore, we will consider the simplest non-trivial LIOM in the main text where $J_{2,1} \neq 0$ and $J_{2,r} =0$ for all $r>1$.

\begin{figure}[t!]
    \includegraphics[width=0.95\columnwidth]{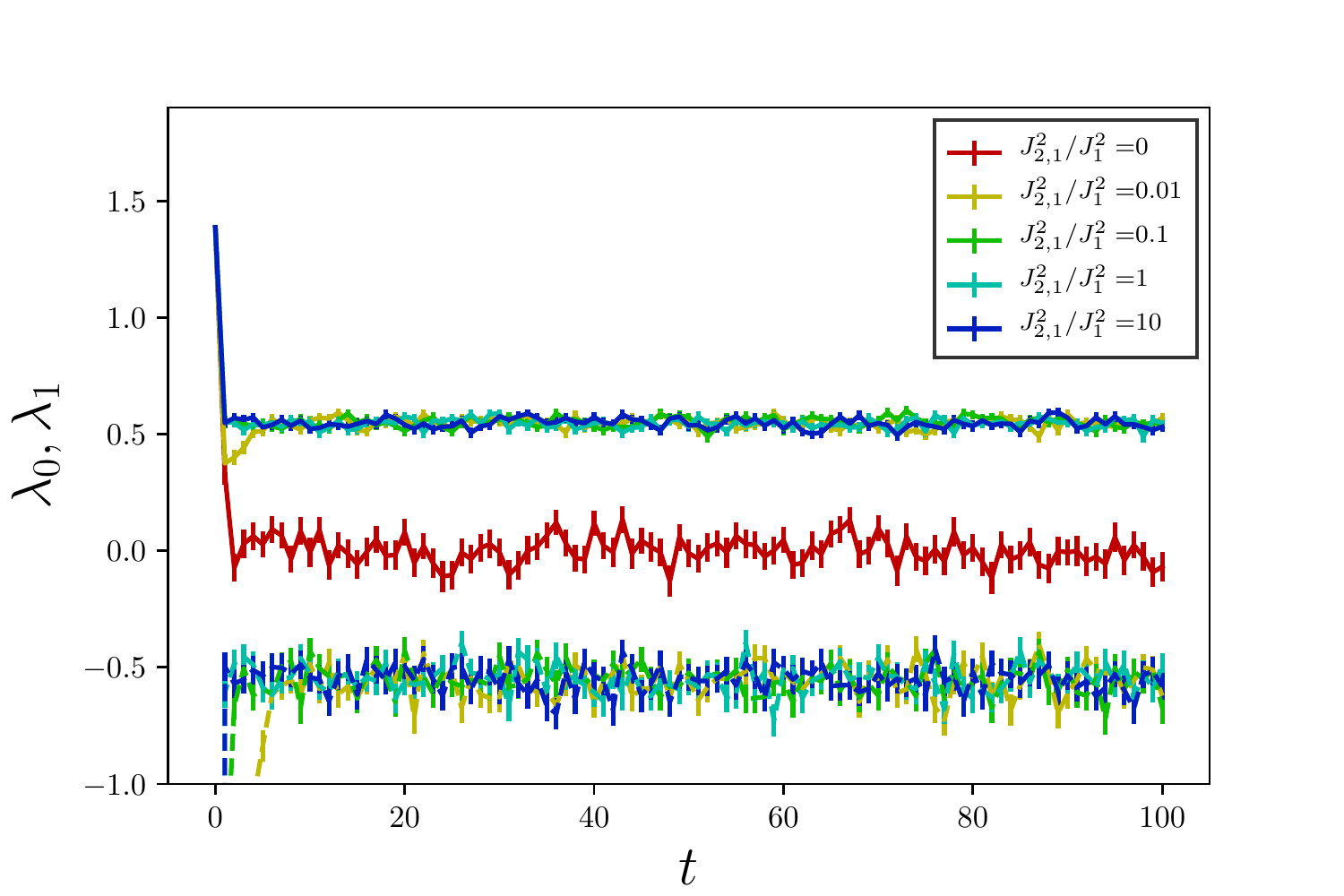}
	\caption{
	The leading (solid lines) and sub-leading (dashed lines) Lyapunov exponents vs time for different ratios between the variances $J_{2,1}^2$ and $J_1^2$. For any finite ratio $J_{2,1}^2 / J_1^2 >0$, the leading exponent is finite at sufficiently large $t$. Note that the sub-leading Lyapunov at $J_{2,1}^2 / J_1^2=0$ converges to a large negative number and is not shown in the plot.
	} \label{fig:liom_lyap}
\end{figure}

To analyze the behaviour of Lyapunov exponents in LIOM, we construct a $2\times2$ transfer matrix for all time $t$, 
\be \label{eq:liom_transfer}
V_i = 
\begin{pmatrix}
e^{-\imath t \left(J^{(1)}_{i+1} +  J^{(2)}_{i,i+1} \right) } & 
e^{-\imath t \left(J^{(1)}_{i+1} - J^{(2)}_{i,i+1} \right) }
\\
e^{-\imath t \left(- J^{(1)}_{i+1} - J^{(2)}_{i,i+1}\right) } & 
e^{-\imath t \left(J^{(1)}_{i+1} + J^{(2)}_{i,i+1}\right) } 
\end{pmatrix}\;,
\ee
such that  
$\Tr_{\mathcal{\tilde H}} [V^{(L)}]  =  \Tr_{\mathcal{\tilde H}} [V_L V_{L-1}\dots V_1]$.
We numerically compute the two Lyapunov exponents using the method of QR decomposition described in Sec.~\ref{sec:numerics}. We see that for any finite ratio $J_{2,1}^2 / J_1^2 >0$, the leading exponent $ \lambda_>\equiv \lambda_0$ and sub-leading exponent $\lambda_1$ converge to positive and negative finite values respectively, as shown in Fig.~\ref{fig:liom_lyap}.

Moreover, we analyze $\langle \mathcal{K}(t)^\alpha \rangle$ for the LIOM \eqref{eq:liom} with two-body terms for integer $\alpha$ in Appendix \ref{app:mbl_hpsff}. We map $\langle \mathcal{K}(t)^\alpha \rangle$ to the partition function of stacked spin chains with two-body interactions, which can be written in terms of another transfer matrix, whose size increases as $\alpha$ increases and as we include longer range two-body terms in \eqref{eq:j1j2}. 
We numerically diagonalize the transfer matrix and show 
that the $F_t(\alpha)$ for integer $\alpha$ are qualitatively consistent with
the Lyapunov exponents calculation above, and with the form of $F_t(\alpha)$ computed for the RPM and KIM in MBL regime, as discussed below in Sec.~\ref{sec:numerics}.

We have used the LIOM picture to show that the leading Lyapunov exponent converges to a finite value as a function of time. We expect the existence of a positive finite $\lmax$ to persist for general LIOM with exponentially decaying support (see examples in Appendix \ref{app:mbl_hpsff}) and higher-body interaction terms. 
As one includes interaction terms of larger supports in the analysis, the size of the transfer matrix  \eqref{eq:liom_transfer} and, consequently, the number of Lyapunov exponents increases. 
However, intriguingly, there is not a notion of time-momentum sectors for the Hamiltonian in Eq.~\eqref{eq:liom} once expressed in the LIOM basis. This seems to indicate the possibility of a further structure for the LIOM effective Hamiltonian which would retain the notion of a time-momentum quantum number. We will leave the analysis of Lyapunov exponents for Hamiltonian systems for future studies.

\section{Numerics}\label{sec:numerics}
The advantage of the dual formulation is that the Lyapunov exponents can be computed efficiently via an iterative procedure at arbitrarily large $L$. Indeed, by using the QR decomposition, we can write
\begin{align}\label{eq:QR}
\begin{split}
V_1 &= Q_1 R_1  \; ,
\\ 
V_2 Q_1 &= Q_2 R_2 \; ,
\\
V_L  \dots V_2 V_1 &= Q_L R_L R_{L-1}\dots R_1 \; ,
\end{split}
\end{align}
where $Q_i$ is an orthogonal matrix and $R_i$ is an upper triangular matrix. An estimate of the $a$-th Lyapunov exponent in the momentum sector $k$ is then 
\begin{equation}
\label{eq:lambdamu}
\lambda_{a}(k) = \frac{2}{L}\sum_{i=1}^L  \mu_{a,i}(k) \;,
\end{equation}
where we define for convenience
\begin{equation}\label{eq:mu_i}
\mu_{a,i}(k) \equiv   \ln [R_{i}]_{aa} \;.
\end{equation}

By iteratively acting with the matrices $V_i$ and projecting onto the momentum sector $k$, we can generate a large number $L$ of $\mu_{a,i}$. In this way we can obtain the behavior of $F_t(\alpha)$ for $\alpha$ in the neighbourhood of $0$. However, in order to access larger values of $\alpha \gtrsim 1$, it is necessary to access values of $\lambda^{(L)}_{a} \neq \langle \lambda^{(L)}_a \rangle$ whose probability is exponentially suppressed in $L$. This requires repeating the calculation in Eq.~\eqref{eq:lambdamu} several times in order to sample the tail of the distribution of $\lambda^{(L)}_a$ at finite $L$. 
To this end, we define 
\be  \label{eq:mu_def_ell}
\lmax^{(\ell)} := \frac{2}{\ell} \sum_{j=1}^{\ell} \mu_{0, j}(k=0) \; ,
\ee
where $\ell$ is chosen such that the spatial correlation between $\mu_{0}$ and $\mu_{\ell}$ is sufficiently small. Our data suggest that, for both the RPM and KIM simulations, it is sufficient to have $\ell = 10$, which we will take hereafter. We then define an effective cumulant generating function that approximates Eq.~\eqref{eq:Fdef} as
\be \label{eq:Fdef_approx}
 F_{t,\ell}(\alpha) := \frac{1}{\ell} \log \langle 
e^{2 \alpha \ell \lmax^{(\ell)}} \rangle \;.
\ee
where $\langle \cdot \rangle$ denotes the average over all realizations of $\ell$ consecutive $\mu$'s in \eqref{eq:mu_def_ell}. 

We can perform this numerical procedure exactly and the main limitation is represented by the exponential growth in the size of the matrices $V_i$ with $t$. Alternatively, one can adopt some approximate scheme based on matrix-product states (MPS) and the density-matrix renormalization group (DMRG) algorithm. However, we will see below that this is effective only deep in the MBL phase. 

Using these methods, we compute the leading Lyapunov spectrum $\lambda_0 (k)$, focusing in particular on two main representative cases  $\lmax\equiv \lambda_0(k=0)$ and $\lambda_0(k=\pi)$ as functions of time $t$. We also extract the cumulant generating function $F_{t,\ell}(\alpha)$ in the chaotic and MBL phases. 
At late time in the MBL phase, we expect $\lambda_0 (k)$ to have a non-uniform shape as a function of $k$ with a positive finite $\lmax$ in the $k=0$ momentum sector. 
In the chaotic phase, we expect the leading Lyapunov $\lambda_0 (k=0)$ to approach zero at late time, and we further conjecture that the largest Lyapunov exponents in the other momentum sector approach zero as well, so that $\lambda_0 (k)$ is flat in the chaotic phase.
%
%
Finally, we expect $F_{t,\ell}(\alpha)$ to have a finite positive gradient in the MBL phase, and to have zero gradient in the chaotic phase.
We summarize the result of numerics as follows: For the KIM, the data are in agreement with the theoretical expectations above.  
Note that, exactly at the self-dual point of the KIM, $V_{i}$ is unitary. Consequently, the SFF does not grow exponentially in space, and $ \lambda_>$  is identically zero at the self-dual point. For this reason, even away from the self-dual point, the finite-time corrections are small.
For RPM with on-site dimension $q=3$, the data are compatible with the theoretical expectations, but agreement is not conclusive due to the limited times that are accessible within our numerics.

	\begin{figure}[t!]
		\includegraphics[width=1\columnwidth]{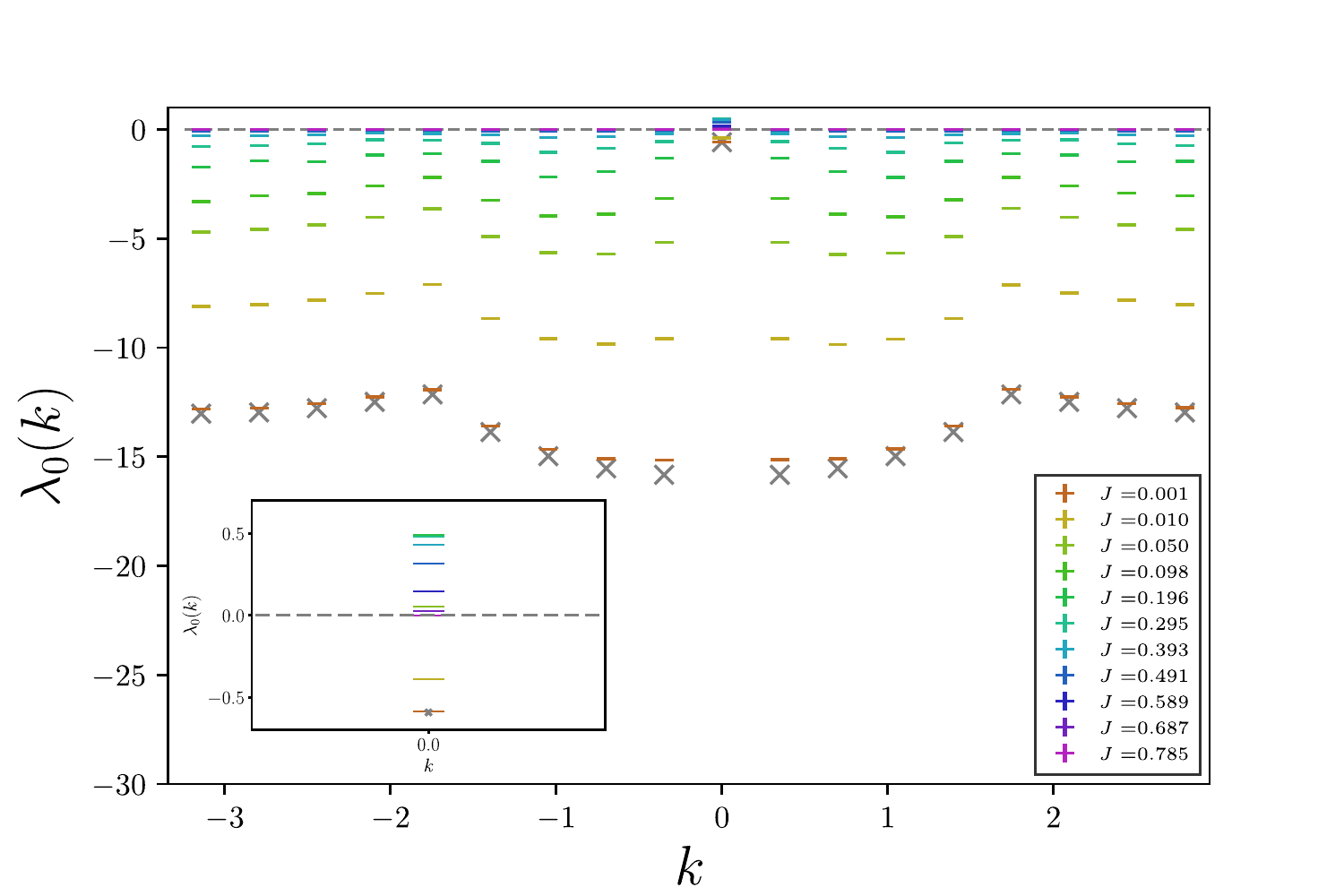}
		\caption{
		Main panel:
		$ \lambda_0(k)$ vs $k$ for KIM with $t=18$ for a range of values of $J$. (Recall that the critical coupling strength for KIM is $J_c= 0.23$). We include data for $ \lambda_0(k) $ at values of $J$ as small as $0.001$ and compare it with the result from perturbation theory, Eq.~\eqref{eq:pertfinal}, labelled in grey (full equation in \eqref{eq:perturb_full_kim}.)
		Inset: 	$\lambda_0(k=0)$ for different values of $J$ with the same colour coding as the main panel.  
		} \label{fig:lam_vs_k_kim}
	\end{figure}	
\begin{figure}[t!]
		\includegraphics[width=1\columnwidth]{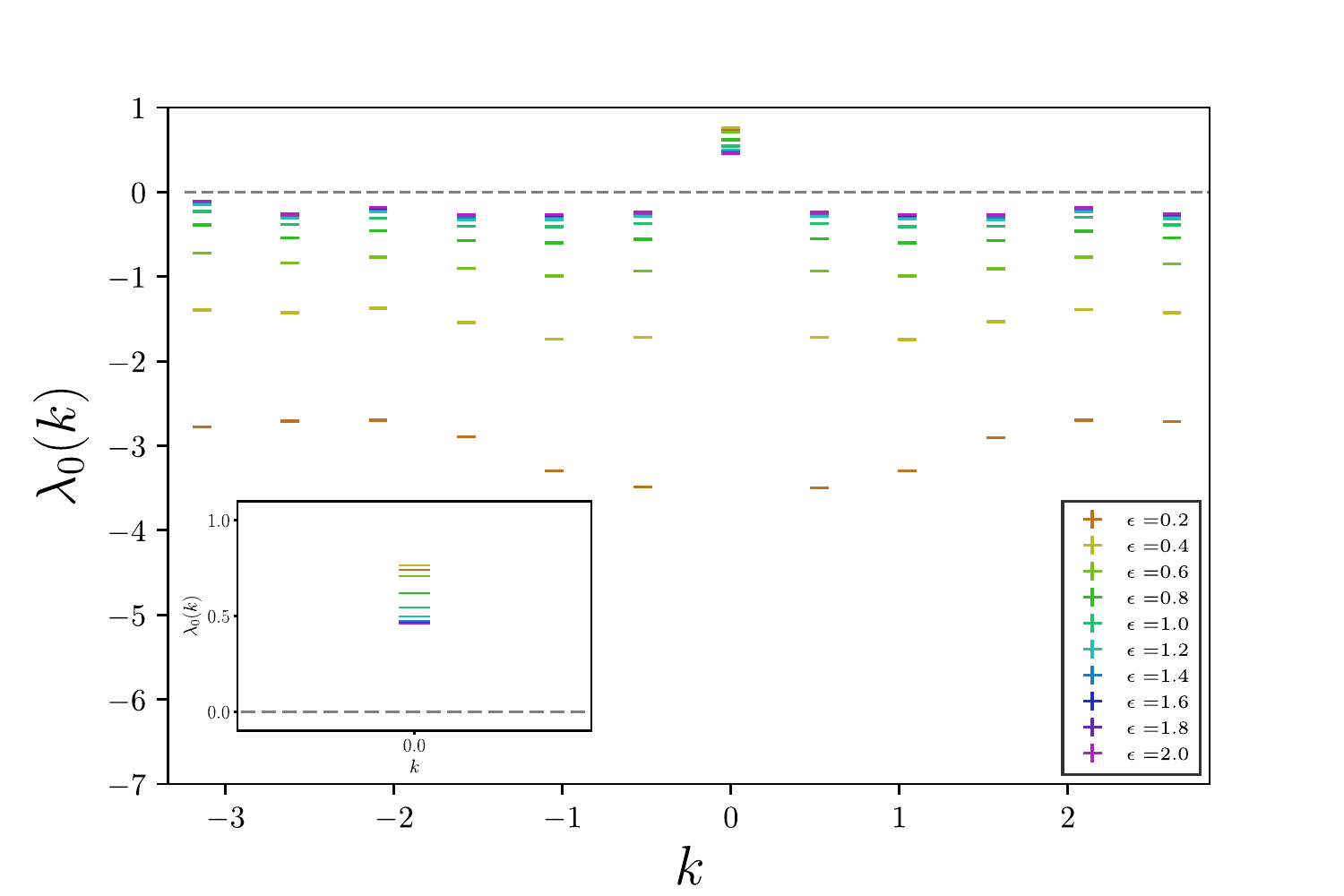}
		\caption{
		Main panel:
		$ \lambda_0(k) $ vs $k$ for RPM with $t=12$ for a range of values of $\epsilon$. 
		(Recall that the RPM at $q=3$ has  $\epsilon_c\approx 0.25$.)
		Inset: $\lambda_0(k=0)$ for different values of $J$ with the same colour coding as the main panel.
} \label{fig:lam_vs_k_rpm}
	\end{figure}

In Fig.~\ref{fig:lam_vs_k_kim} and~\ref{fig:lam_vs_k_rpm}, we show the largest Lyapunov exponents $ \lambda_0(k) $ in each momentum sector $k$ for the KIM and RPM respectively. 
For the KIM in the chaotic phase, $\lambda_0(k)$ is very small for all $k$. 
On the other hand, in the MBL phase, $ \lambda_0(k=0) $ is positive (except for very small $J$, see below), and $ \lambda_0(k\not=0) $ is negative.
In Fig.~\ref{fig:lam_vs_k_kim} we include data for $J$ as small as $0.001$ and show that, for $k\neq 0$, it agrees well with the result from perturbation theory given in Eq.~\eqref{eq:pertfinal} (full equation in \eqref{eq:perturb_full_kim}), and that for $k=0$ it agrees with the result from degenerate perturbation theory evaluated numerically.
Note that in Fig.~\ref{fig:lam_vs_k_kim} we observe a peculiarity in $\lambda_0(k=0)$ for $J=0.001, 0.01$, where the $\lambda_0(k=0)$ have small negative values. We find that the window of $J$ where $\lambda_0(k=0)<0$ gets smaller as $t$ gets larger and we expect this to be only a finite-time effect.
For RPM, the data shown in Fig.~\ref{fig:lam_vs_k_rpm} are  limited by finite-$t$ effects, but they are compatible with and seem to tend towards the expected behaviours. 

%

\begin{figure}[t!]
	\includegraphics[width=0.95\columnwidth]{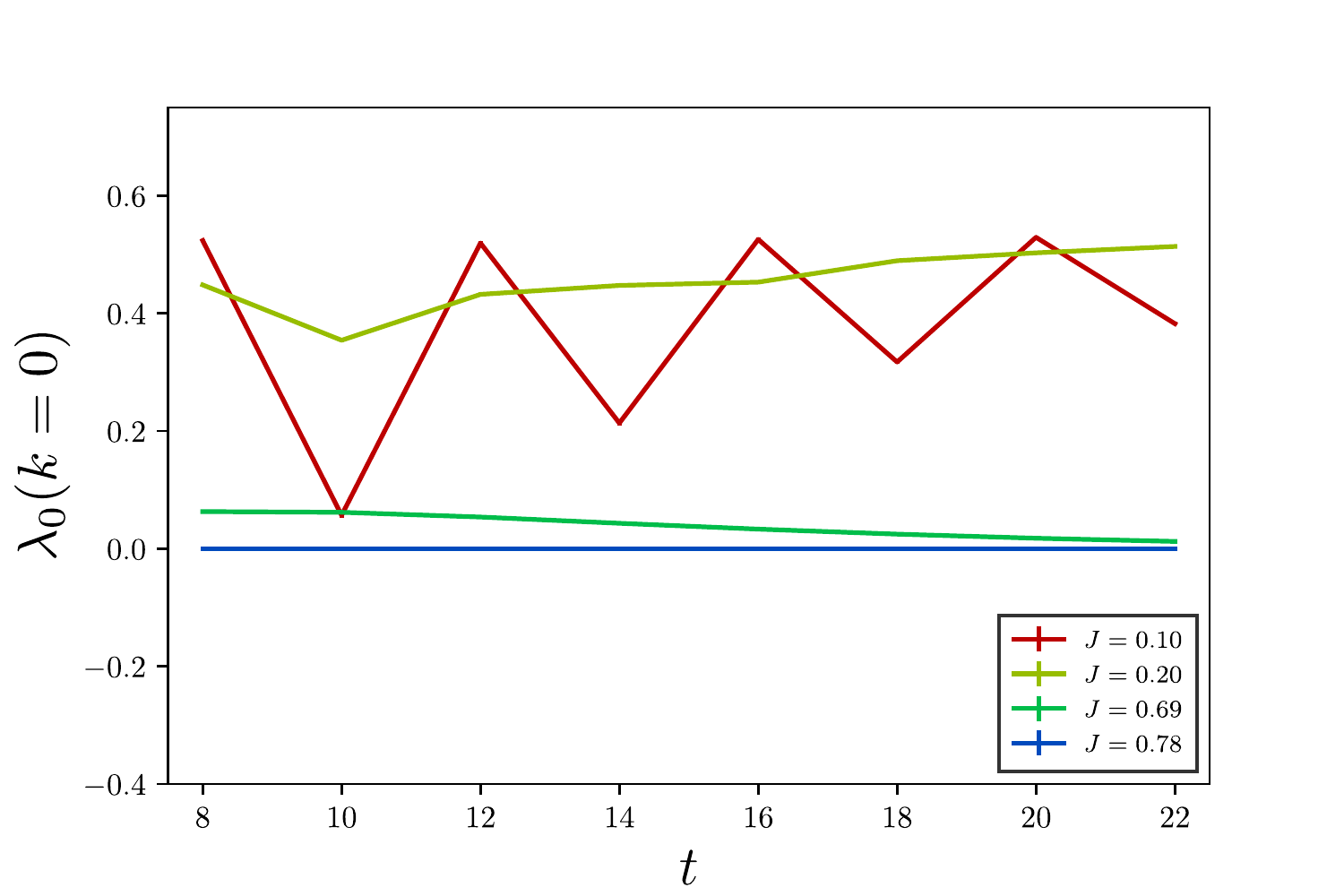}
	\caption{$ \lambda_>$  vs $t$ for KIM  for different values of $J$. 
	} \label{fig:lambda_v_t_kim}
\end{figure}

\begin{figure}[t!]
	\includegraphics[width=0.95\columnwidth]{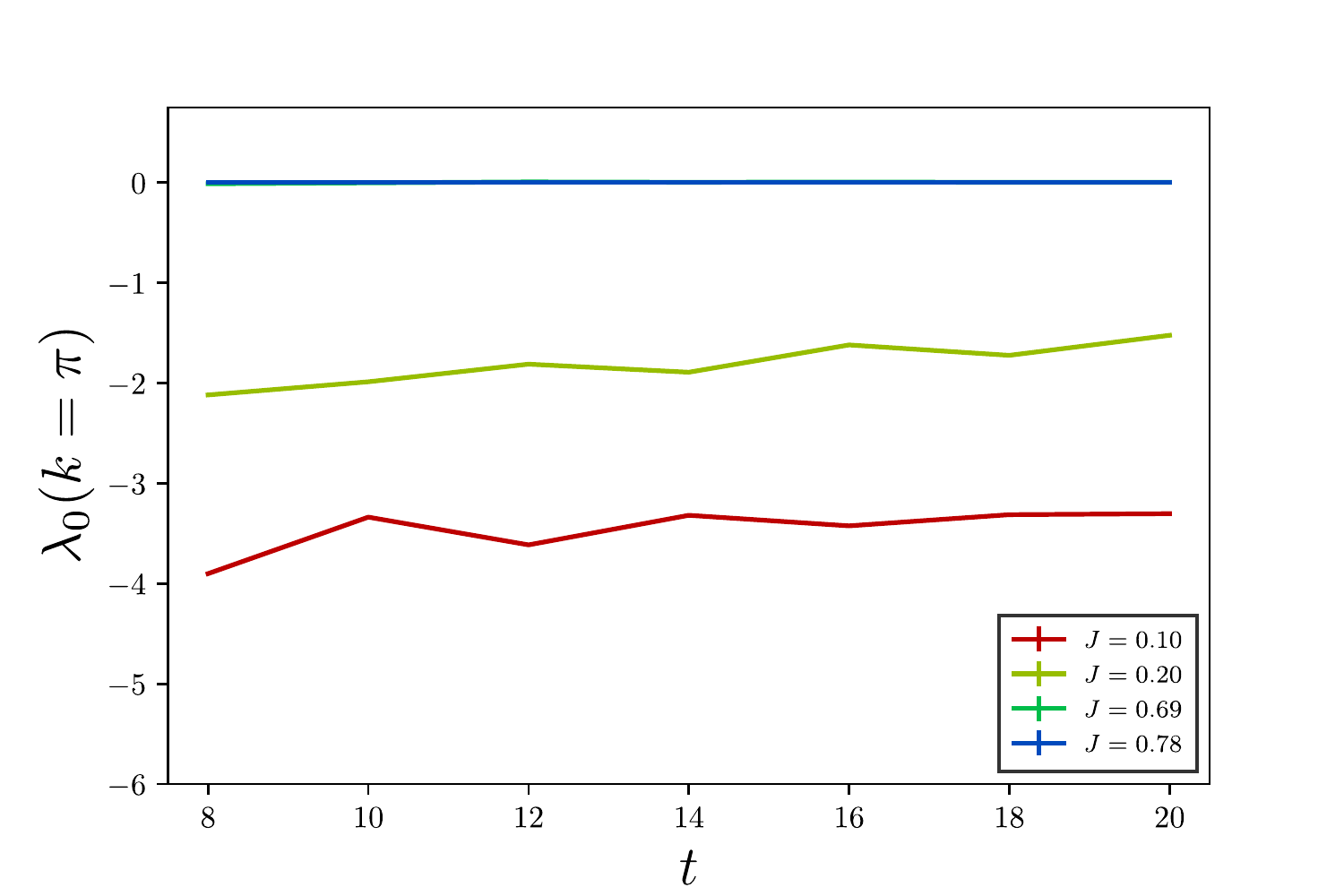}
	\caption{$ \lambda_0(k=\pi) $ vs $t$ up to $t=20$ for KIM with four different values of $J$. 
		Note that the data for $J=0.69$ and $J=0.78$ lie on top of each other.
	} \label{fig:lambda_v_t_k_pi_kim}
\end{figure}

\begin{figure}[t!]
	\includegraphics[width=0.95\columnwidth]{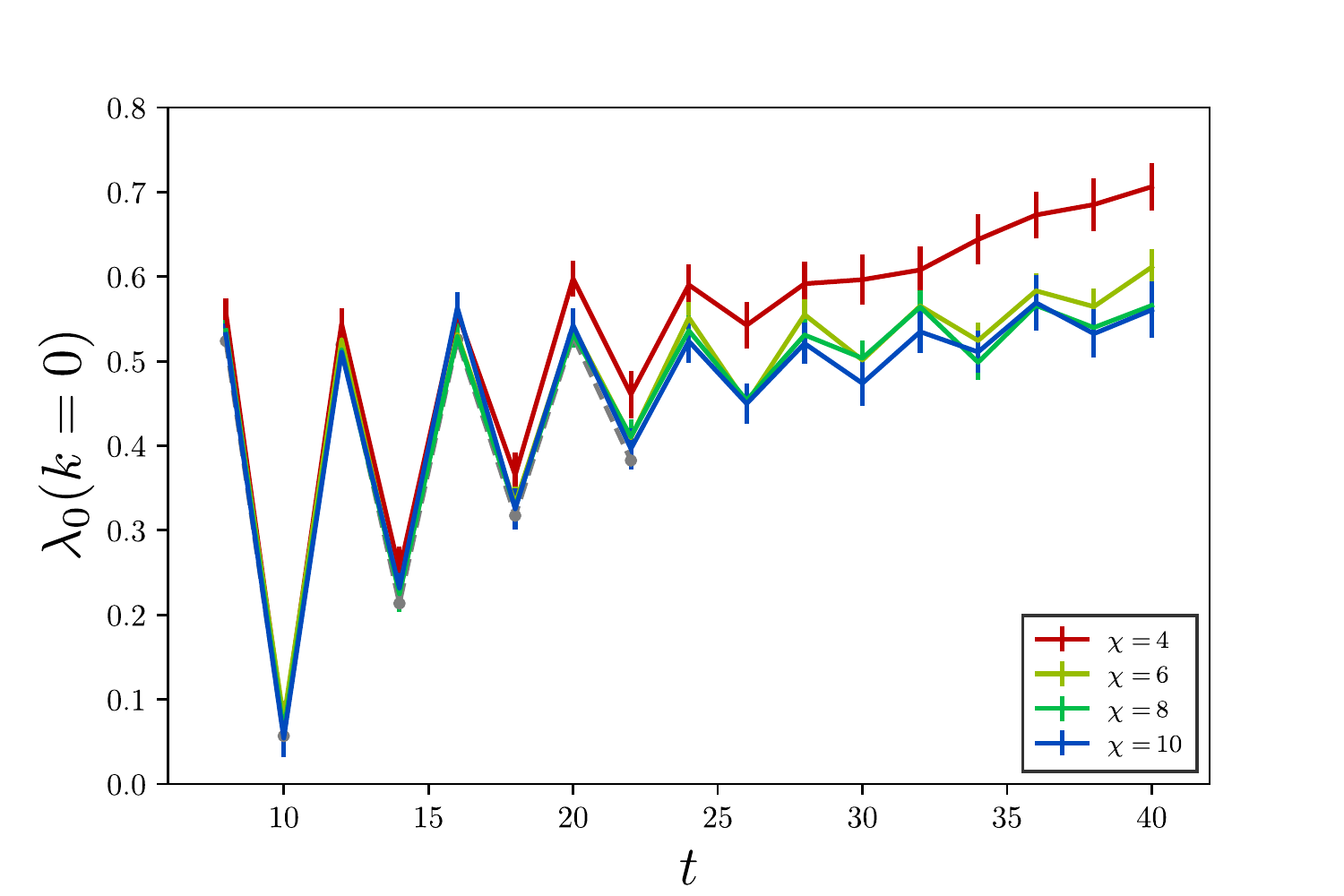}
	\caption{$ \lambda_> $ vs $t$ for KIM  for $J=0.1$ with data from MPS projection on bond dimensions $\chi = 4,6,8, 10$. For comparison, data from exact diagonalisation are shown with a dashed line.
	} \label{fig:kim_w_mps}
\end{figure}

Next, in order to characterize the $t$-dependence of the spectral Lyapunov exponents, we focus on two distinctive cases: $k=0,\pi$. 
In Fig.~\ref{fig:lambda_v_t_kim} and \ref{fig:lambda_v_t_k_pi_kim}, we show  $\lmax\equiv\lambda_0(k=0)$ and $\lambda_0(k=\pi)$ respectively as a function of $t$ for the KIM. Consistently with our picture, in the chaotic phase both $\lambda_0(k=0)$ and $\lambda_0(k=\pi)$ are small at large $t$. In the MBL phase, $ \lambda_0(k=0)$ converges towards a positive value while $\lambda_0(k=\pi) $ tends towards a finite negative value as $t$ increases. Note that there are decaying oscillations in time with a periodicity of $4$ for small $J$ which are still visible at the accessible time with exact matrix multiplication ($t \sim 20$). In order to access larger values of $t$, we employ a variation of the DMRG algorithm: 
after the application of each transfer matrix, we re-project the dual Hilbert space onto a matrix product state at fixed bond dimension $\chi$. 
With this method, we can access much larger times ($t \sim 40$) and confirm that the oscillations are suppressed in $t$, as shown in Fig.~\ref{fig:kim_w_mps}. 
However, the accessible values of $\xi$ are limited by the necessity of using periodic boundary conditions in the time direction, and the non-unitarity of the dual transfer matrix.
In the chaotic phase, the DMRG algorithm applied in the dual picture cannot be exploited for large $t$ since the Lyapunov exponents obtained in this way do not converge for accessible values of $\chi$.

\begin{figure}[t!]
	\includegraphics[width=0.95\columnwidth]{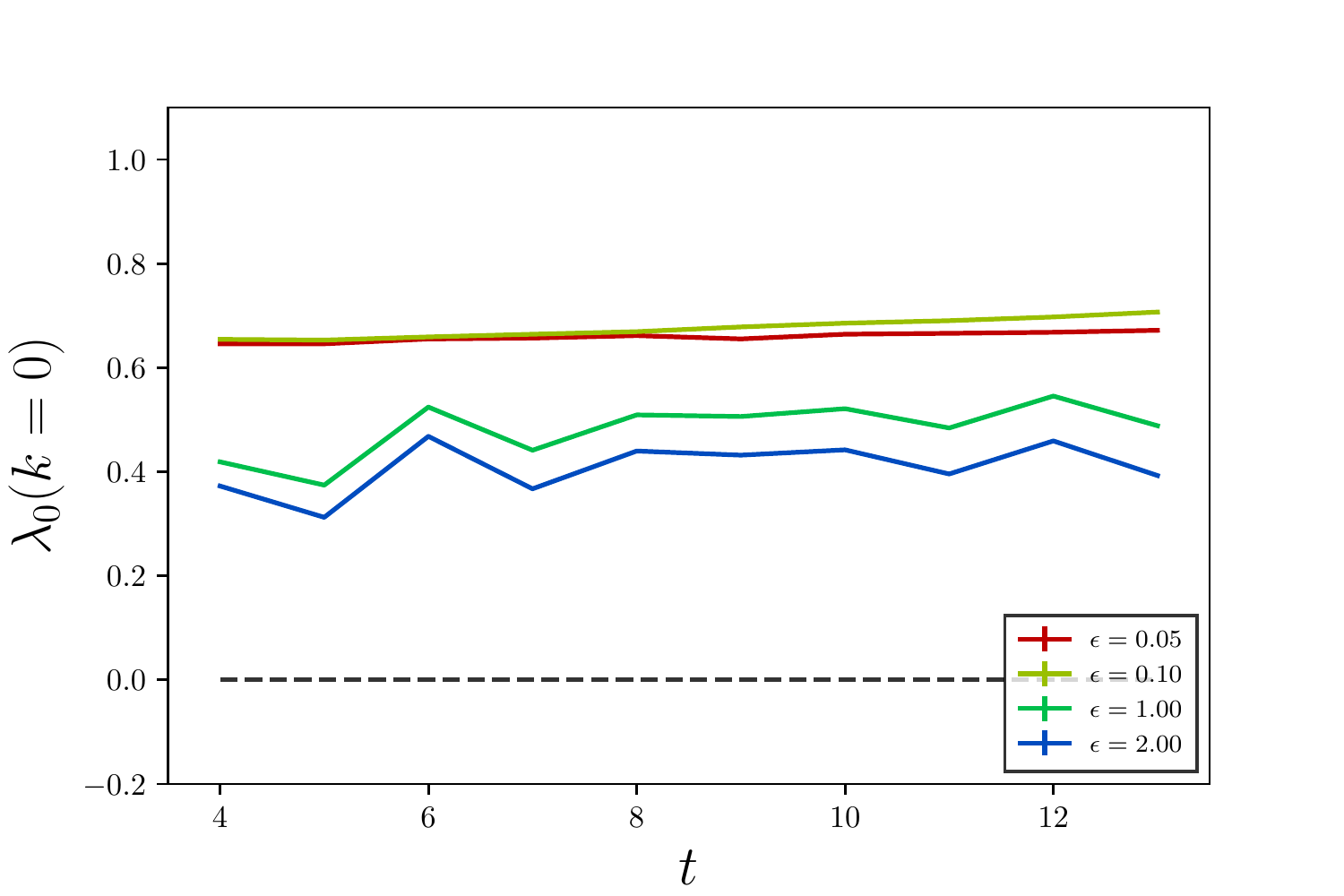}
	\caption{$ \lambda_>$ vs $t$  for RPM for different values of $\epsilon$.
	} \label{fig:lambda_v_t_rpm}
\end{figure}

\begin{figure}[t!]
	\includegraphics[width=0.95\columnwidth]{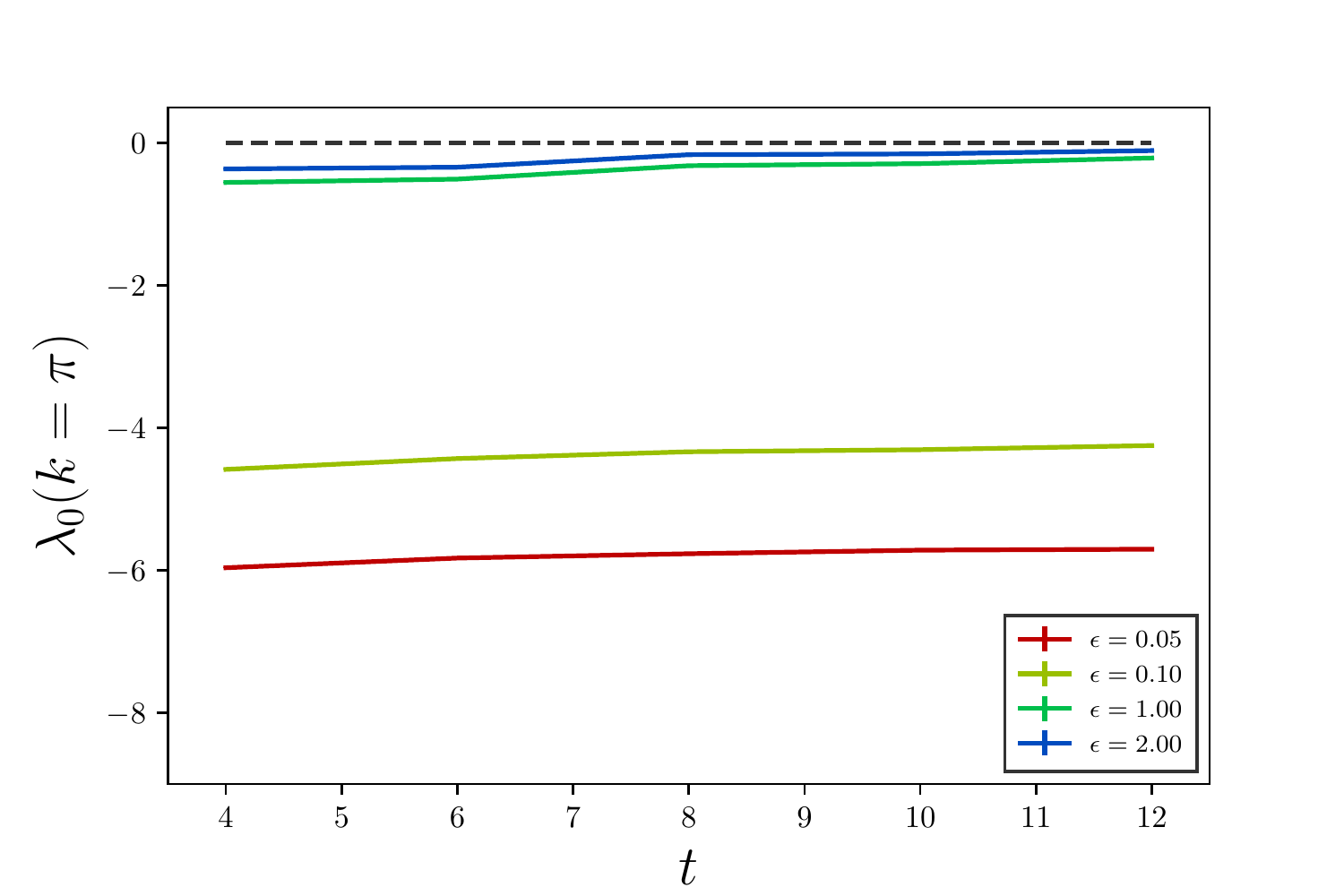}
	\caption{$ \lambda_0(k=\pi) $ vs $t$ for RPM with four different values of $\epsilon$.
	} \label{fig:lambda_v_t_k_pi_rpm}
\end{figure}

In Fig.~\ref{fig:lambda_v_t_rpm} and \ref{fig:lambda_v_t_k_pi_rpm}, we show $ \lambda_0(k)$ against $t$ for the RPM for $k=0$ and $\pi$ respectively. In the MBL phase, $ \lambda_0(k)$ behaves as expected for both momentum sectors. However, the behaviour of $\lambda_0(k)$ in the chaotic phase is affected by the finite time effects. While $ \lambda_0(k=\pi)  $ for the chaotic phase tends towards zero and is small relative to the corresponding Lyapunov exponents in the MBL phase, $ \lambda_0(k=0) $ remains finite for the accessible values of $t$.

\begin{figure}[t!]
	\includegraphics[width=0.95\columnwidth]{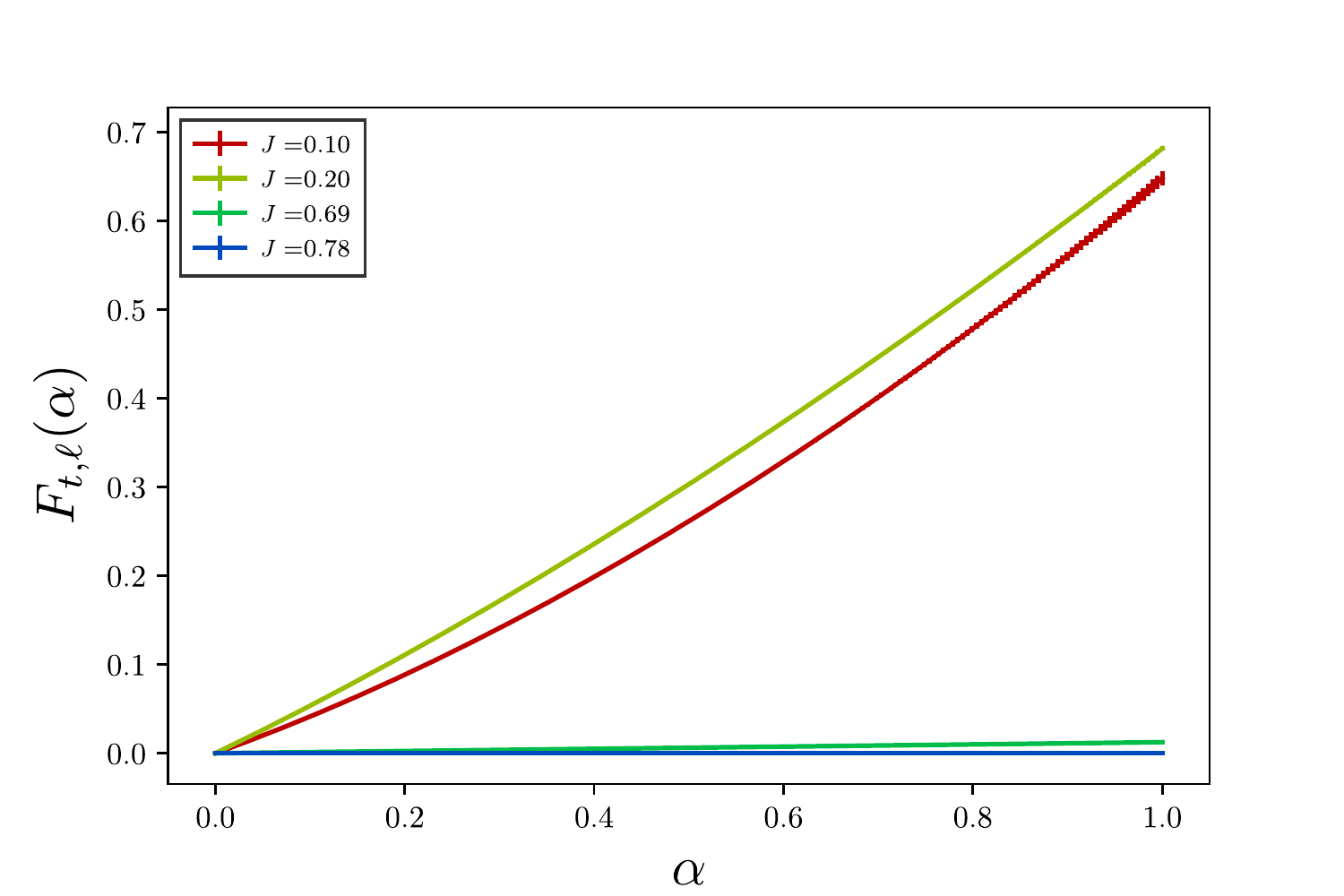}
	\caption{$F_{t,\ell}(\alpha)$ vs $\alpha$ for the KIM with $t= 22$ and $\ell = 10$ for four values of $J$ inside the MBL and chaotic phases. 	
	} 
\label{fig:f_vs_alpha_kim}
\end{figure}

\begin{figure}[t!]
	\includegraphics[width=0.95\columnwidth]{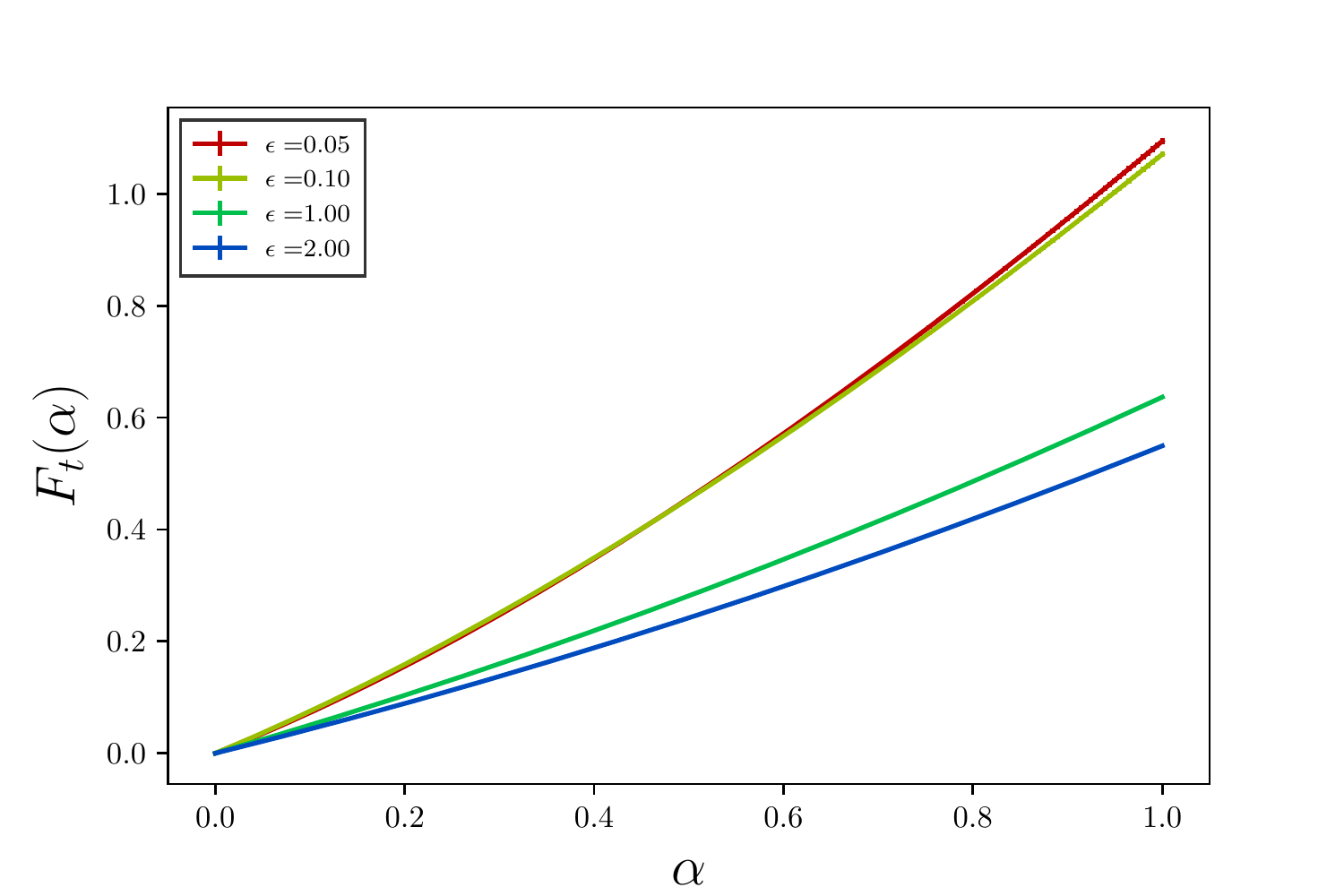}
	\caption{
		$F_{t,\ell}(\alpha)$ vs $\alpha$ for the RPM with $t= 13$ and $\ell = 10$ for four values of $\epsilon$ inside the MBL and chaotic phases.
	} \label{fig:f_vs_alpha_rpm}
\end{figure}

In Fig.~\ref{fig:f_vs_alpha_kim} and \ref{fig:f_vs_alpha_rpm}, we show the cumulant generating function \eqref{eq:Fdef_approx} computed for the KIM and RPM respectively. Recall that the first and second cumulants of $ \lambda_>$ are the first and second derivatives of the cumulant generating function at $\alpha=0$.
For the KIM, $F_{t,\ell}(\alpha)$ shows obviously distinctive behaviours in the chaotic and MBL phases. In particular, $F_{t,\ell}(\alpha)$ has zero derivative in the former phase, which is consistent with the expectation that $ \lambda_>  =0 $, discussed in earlier sections. However, again, for RPM, $F_{t,\ell}(\alpha)$  does not show such a clear difference in behaviour between the two phases for the accessible $t$ (Fig.~\ref{fig:f_vs_alpha_rpm}).

Finally, we recall the different symmetry classes of the KIM and RPM, namely COE  and CUE respectively. The former symmetry class has $K_{\mathrm{COE}} \approx 2t$. Therefore, for the KIM, it is natural to expect in the chaotic phase at large times that there are $2t$ (not just $t$) zero Lyapunov exponents contributing to $ K(t) \sim \sum_{k,a} e^{\lambda_a(k) L}$, two from each of the $t$ momentum sector. In order to check this, we compute the gaps $\Delta \lambda_a(k)\equiv \lambda_a(k) -\lambda_{a+1}(k)$ in Appendix.~\ref{app:lyap_gap}, and verify that $\Delta \lambda_0(k)$ is indeed small {at large $t$} in the chaotic phase for the KIM. In RPM, the corresponding computation shows that the gap $\Delta \lambda_a(k)$ is much larger.

\section{Concluding remarks}\label{sec:dis}

We have proposed a new set of physical quantities, the spectral Lyapunov exponents, which allow us to explore the fluctuations and the generic behaviour of the SFF in the thermodynamic limit. We have shown that the spectral Lyapunov exponents have distinct long-time behaviours in the chaotic and MBL phases: For chaotic systems, the largest Lyapunov exponent in each momentum sector $k$ converges to zero at large time, implying the absence of exponential growth of $K(t)$ with system size and the onset of random matrix behavior in the spectral correlation. For MBL systems, the Lyapunov exponents remain non-zero with a non-universal form of the spectrum which encodes the residual spectral correlations. 
We further propose a scaled cumulant generating function $F_{t}(\alpha)$ associated with the hpSFF, which encodes the fluctuations of the leading Lyapunov exponent in the zero-momentum sector. We argue on the basis of analytical and numerical analyses that the average $F'_t(0) = \lim_{L\to\infty} L^{-1}\langle \ln K(t)  \rangle$ provides a sufficient characterization of the MBL / chaotic phase in generic settings.

Our results for behaviour of the spectral Lyapunov exponents in each phase are complementary to and consistent with recent studies based on a transfer matrix that generates the average SFF \cite{garratt2020manybody,garratt2020MBL}.

There are many interesting directions to pursue in the future. First, it would be exciting to look at the behavior of spectral Lyapunov spectrum when the MBL-ETH transition is approached and where universality is expected and could manifest itself both in the fluctuations $F_t(\alpha)$ and the spectrum $\lambda_0(k)$. 
Second, it remains to understand how the existence of conserved quantities affects the behavior of the spectral Lyapunov exponents. One possible extension would be the inclusion of a $U(1)$ charge conservation~\cite{friedman2019}. More generally, one could look at the behavior of Hamiltonian systems for which the energy provides a natural conserved quantity. In such cases, the time variable in the dual picture is continuous and the time momentum operator becomes a local conserved quantity in contrast to the Floquet case. This should be at the origin of the different scaling expected for the Thouless time in these systems.

\section{Acknowledgement}
AC is supported by fellowships from the Croucher foundation and the PCTS at Princeton University. JTC  is  supported  in  part  by  EPSRC  Grants  EP/N01930X/1  and EP/S020527/1.

\bibliography{FloquetChaos}

\onecolumngrid
\appendix

\section{Derivation of $ \lambda_> $ and $F_t(\alpha)$ in chaotic phase} \label{app:F_eth}
In this Appendix, we compute $\langle \mathcal{K}(t)^\alpha \rangle$ for the RPM in the limit of large $q$ and large $L$ by obtaining the leading eigenvalue of the transfer matrix \eqref{eq:rp_tmat}. Furthermore, we analytically continue the results to compute $F_t(\alpha)$ and $\lambda_>(t)$ in the same limits. 

To obtain the leading eigenvector of the transfer matrix \eqref{eq:rp_tmat} with integer $\alpha = n$, note that all of its matrix elements are non-negative. So there is a unique largest real eigenvalue and a corresponding eigenvector with non-negative components due to the Perron-Frobenius theorem. Furthermore, due to the symmetry of the diagrams, the eigenvector must be invariant under permutation, and hence we find $(1,\dots, 1)^T$ as the leading eigenvector. 

To find the leading eigenvalue $E_1$, we sum over any given  row of $T$, and obtain
\begin{subequations}\label{eq:leading_evalue}
	\begin{align}
	&E_1 = \left(\frac{t}{1 + t y}\right)^n\sum_{d=0}^\thetaint P(\thetaint, d) \, (1 + y)^{n-d}
	\label{leadeigen}
	\\
	&P(\thetaint, d ) = \frac{\thetaint !}{(\thetaint - d)!} \sum_{j=0}^d \frac{(-1)^j}{j!}
	\label{Pterm}
	\\
	&y = \frac{e^{t \epsilon} -1}{t}
	\end{align}
\end{subequations}
where $P(\thetaint, d)$ is the number of elements in $S_\thetaint$ with distance $d$ from any given reference permutation\cite{diaconis1988}, say the identity $\mathbf{p}=(1,2,\dots, \thetaint)$; $C(t,\epsilon, \thetaint, d) = \sum_{\mathbf{v'}} \bra{(1,\dots, n), (1,\dots, 1)} T \ket{\mathbf{p}, \mathbf{v}'}$ is the sum of $t^\thetaint$ matrix elements at fixed $\mathbf{p}$. 
From the leading eigenvalue in \eqref{leadeigen}, we can then recover  $F_t(\thetaint) = \log E_1$.

Above we derived an expression for  $\langle \mathcal{K}(t)^\alpha \rangle$  at integer $\alpha = \thetaint$. Now we re-express Eq.~\eqref{eq:leading_evalue} in a different form where the dependence on $\alpha$ can be easily analytically continued to real values. First of all, we can rewrite the sum in \eqref{Pterm} as
\begin{equation}
\label{partialsum}
\sum_{j=0}^d \frac{(-1)^j}{j!} = \frac1e - \sum_{k=0}^\infty \frac{(-1)^{k+d+1}}{\Gamma(k + d + 2)}
\end{equation}
where $e$ is the Neper number. 
Plugging \eqref{partialsum} in \eqref{Pterm}, we can exchange the order of sums in \eqref{leadeigen} and perform the sum over $d$. After some manipulations, the final result takes the compact form valid for arbitrary $\alpha \geq 0$ 
\begin{equation}
E_1 = e^y \left(\frac{1}{t}+y\right)^{-\alpha } \Gamma (\alpha +1,y) \;,
\end{equation}
and 
\begin{equation}
\lim_{q\to \infty} F_t(\alpha) = \log E_1 \;.
\end{equation}

Leaving the large-$q$ limit implicit, we can now compute the leading Lyapunov exponent
\begin{equation}
 \lambda_>  = F_t'(\alpha = 0) = \lim_{L \to \infty}\frac{1}{L} 
\langle \log \mathcal{K}(t)\rangle
=
\lim_{L \to \infty}\frac{1}{L} 
\frac{\partial }{ \partial\alpha } 
\left.
\langle \mathcal{K}(t)^\alpha\rangle 
\right|_{\alpha \to 0} \; .
\end{equation}
The derivative of the incomplete Gamma function can be evaluated as
\begin{equation}
\left. \frac{\partial }{ \partial \alpha }  \Gamma (\alpha+1,y) \right|_{\alpha \to 0} = e^{-y} \log (y)+ \Gamma(0,y) \;.
\end{equation}
After some straightforward manipulations, we arrive Eq.~\eqref{eq:lambda_rpm}, reproduced below,
\begin{equation}
\lambda_> =
e^y\Gamma(0,y)
-\log \left(1+ \frac{1}{yt}\right) \;. 
\end{equation}

\section{Explicit form of the dual circuit \label{app:dual}}
Here, we derive an explicit form for the dual transfer matrix for the two models introduced in Sec.~\ref{sec:models}. 
Both models are composed of a layer $W_1$ of single-site unitaries and a layer $W_2$ of $2$-site unitaries diagonal in the computational basis. 
We will therefore treat them both at once. To be more specific, we use the notation introduced in Sec.~\ref{subsec:RPM} for the RPM in Eqs.~(\ref{eq:W1}, \ref{eq:W2}), i.e.
\begin{subequations}
\begin{align}
    &W_1 = U_1 \otimes U_2 \ldots U_L 
    \;,
    \\
    &[W_2]_{a_1,\ldots,a_L; a_1',\ldots a_L'} = \delta_{a_1,a_1'}\ldots \delta_{a_L,a_L'} \exp\left(\imath \sum_n \varphi_{a_n, a_n+1}^{(n)}\right) \qquad a_n \in {1,\ldots q}
    \;.
\end{align}
\end{subequations}
For the RPM, the unitary matrices $U_j$ are drawn from the CUE and the phases $\phi_{a_n, a_{n+1}}$ are Gaussian variables with zero average and standard deviation $\epsilon$.
With the same notation, the KIM can be recovered  setting $q=2$, with 
$U_j = e^{\imath h_j \sigma_j^z} e^{\imath b \sigma_j^x}$ and 
$\varphi_{a_n, a_{n+1}}^{(n)} =J e^{\imath \pi (a_n + a_{n+1})}$ ($a_n = 1,2$). 

In order to deduce the form of the transfer matrix in the space direction we write explicitly the trace in \eqref{eq:dual}. We introduce a compact notation for the indices $\av = (a_1,\ldots, a_L)$ and we have
\begin{equation}
\Tr_{\mathcal{H}}[W(t)] = \sum_{\{\av^{1},\ldots, \av^{t}\}}[\TT]_{\av^1, \av^t} \ldots [\TT]_{\av^3, \av^2}  [\TT]_{\av^2, \av^1}=
\sum_{\{\av^{1},\ldots, \av^{t}\}} \prod_{j=1}^L \prod_{\mu = 1}^t 
 e^{\imath \varphi_{a_j^\mu, a_{j+1}^\mu}^{(j)} }
[ U_{j}]_{a_j^{\mu+1},a_j^{\mu}} \;.
\end{equation}
We now introduce a dual Hilbert space $\tilde{\mathcal{H}} = \otimes_{\mu=1}^t \mathbb{C}^q$ with dimension $\tilde {\mathcal{N}} = q^t$ and the computational basis 
$\bv = \{b^1, \ldots, b^t\}$ with each $b^\mu= 1,\ldots,q$. Then, defining the $j$-dependent dual layers
\begin{subequations}
\label{eq:VVsdef}
\begin{align}
&[\VV_{1,j}]_{\bv,\bv'} = \prod_{\mu=1}^t e^{\imath \varphi_{b^\mu, {b^\mu}'}^{(j )} } \label{eq:VV1}  \;, \\
&[\VV_{2,j}]_{\bv,\bv'} =  \prod_{\mu=1}^t [ U_j]_{b^{\mu+1}, b^{\mu}} \delta_{\bv,\bv'} 
\;,
\end{align}
\end{subequations}
and $\VV_{j} = \VV_{2,j} \VV_{1,j}$, we have that
\begin{equation}
\label{dualitytrace}
\Tr_{\mathcal{H}}[\TT^t] = \sum_{\{\bv_{1},\ldots, \bv_{L}\}}[\VV_{1}]_{\bv_1, \bv_2}[\VV_{2}]_{\bv_2, \bv_3} \ldots [\VV_{L}]_{\bv_L, \bv_1}  = \Tr_{\tilde{\mathcal{H}}}
[
\VV_{1} \VV_{2} \ldots \VV_{L}
] \;.
\end{equation}
Note that in the dual formulation the $1$-body unitary matrices in $\TT_1$ are converted into $2$-body diagonal matrices in $\VV_2$, while the $2$-body phases in $\TT_2$ are converted into the $1$-body $\VV_1$.

\section{Weakly coupled spins}
\label{app:pert}
In this Appendix we provide the details of the calculation of the Lyapunov spectrum in the limit where different sites are weakly coupled. This corresponds to $J \to 0$/$\epsilon \to 0$ respectively for the KIM/RPM. For the sake of clarity, we will focus on the KIM, although the discussion can be easily adapted to the RPM. 

From Eqs.~\eqref{eq:VVsdef}, we have
\begin{subequations}
\label{eq:dualspin}
\begin{align}
&\VV_1 \equiv \prod_{\mu=1}^t (e^{\imath J}  {\mathbf{1}}_{\mu} + e^{-\imath J} {\sigma}_\mu^x) \label{V1spin}
= [2 \imath \sin(2 J)]^{t/2} 
e^{\sum_\mu \ffun(J_j) \sigma^x_j } 
\\
&\VV_{2,j} \equiv  2^{-t} e^{\imath h_j \sum_{\mu=1}^t\sigma_\mu^z}\prod_{\mu=1}^t (e^{\imath b} \mathbf{1}_{\mu,\mu+1} + e^{-\imath b} \sigma_\mu^z \sigma_{\mu + 1}^z) =\left[\frac{\imath}{2} \sin(2 b)\right]^{t/2} e^{\imath h_j \sum_{\mu=1}^t \sigma_\mu^z  + \ffun(b) \sigma^z_\mu 
\sigma^z_{\mu+1}}
\end{align}
\end{subequations}
where in the last equalities we used the matrix identity holding for any operator $ O^2 = {\mathbf{1}}$
\begin{equation}
    e^{\imath a} + e^{-\imath a} {O} = [2 \imath \sin(2a)]^{1/2} e^{\ffun(a)  O}
\end{equation}
and $\ffun(a) = \arctanh(e^{-2 \imath a})$.
Setting $\sigma^x \ket{\pm} = \pm \ket{\pm}$, we define
\begin{align}
    & V_1 \ket{\bf{0}} = (2 \cos J )^t \ket{\bf 0} \;, \qquad \ket{\bf 0} \equiv \ket{+ \ldots +} \\
    & V_1 \ket{\mu_1,\ldots,\mu_M} = (\imath \tan J)^{M} (2 \cos J )^t \ket{\mu_1,\ldots,\mu_M} \;, \qquad \ket{\mu_1,\ldots,\mu_M} \equiv \sigma^z_{\mu_1}\ldots \sigma^z_{\mu_M} \ket{\bf 0} \label{eq:spinflipseval}
\end{align}
At small $J$, the largest eigenvalue is associated with the vacuum ferromagnetic state $\ket{\bf 0}$ and spin flips are suppressed with powers of $\tan(J)$. At the leading order in $J$, we can restrict our Hilbert space to a single spin flip ($M = 1$ in \eqref{eq:spinflipseval}).  
In order to compute the trace in Eq.~\eqref{dualitytrace} in this limit, we need the matrix elements of $V_2^{(j)}$ between pairs of single spin-flip states. They can be written explicitly 
by going back to the original time direction as
\begin{equation}
    \braket{\mu | {\VV}_{2,j} | \nu} = 2^{-t} \Tr[\sigma^z U_j^{\ell}\sigma^z U_j^{t-\ell}] \;, \qquad \ell = |\nu - \mu|
\end{equation}
where the trace is performed in the Hilbert space of a single spin. 

Additionally we can make use of the translational invariance in the time direction to decompose the trace in \eqref{dualitytrace} in momentum sectors. We thus define a spin wave with momentum $k$ as
\begin{equation}
    \ket{k} = \frac{1}{\sqrt{t}}\sum_{\mu} e^{\imath \mu k} \ket{\mu} \;, \qquad k = \frac{2\pi n}{t} \;, \quad n = 0,\ldots, t-1
\end{equation}
The trace in the single spin flip of momentum $k \neq 0$ can then be written as
\begin{equation}
    \label{tracek}
    \Tr_{k}[\VV^{(1)} \VV^{(2)}_{1} \ldots \VV^{(1)} \VV^{(2)}_{L}] = (2 \cos J)^{t L} (\imath \tan J)^L
    \braket{k | \VV^{(2)}_{1} | k} \ldots \braket{k | \VV^{(2)}_{L} | k}
\end{equation}
We deduce 
\begin{equation}
    \lambda_0(k) \stackrel{J \ll 1}{\sim } 2t \ln |2\cos J| + 2\ln |\tan J | + \overline{\ln [|\braket{k | V_{2} | k}|^2]}
\end{equation}
Setting 
\begin{equation}
    \theta_h = \arccos(\cos(b) \cos(h)) \;, \qquad 
    \alpha_h = \arccos\left(\frac{\sin(h)\cos(b))}{\sin(\theta)}\right) \;,
\end{equation}
we can rewrite
\begin{equation}
    U_j = e^{\imath \theta_{h_j} \vec n \cdot \vec \sigma_j } \;, \quad   \vec n = \frac{1}{{\sin (\theta )}}(\sin (b) \cos (h),-\sin (b) \sin (h),\cos (b) \sin (h))
\end{equation}
which can be easily diagonalized and we arrive at the final expression
\begin{equation}
 \overline{\ln [|\braket{k | \VV_{2} | k}|]^2} = - t \log 2 +  \int dh P(h)  \ln\left[\frac{\sin(\alpha_h)^2 \sin(2 \theta_h) \sin(t \theta_h)}{\cos(k) - \cos(2\theta_h)}\right]^2 \;.
\end{equation}
At large $t$, we can make the replacement inside the integral $\ln |\sin(t \theta_h)|^2 \to -2 \log 2$ and for $b = \pi/4$, we get the final expression
\begin{equation}\label{eq:perturb_full_kim}
    \lambda_0(k) \stackrel{J \ll 1}{\sim } 2t \ln |\cos J| + 2 \ln | \tan J| + \int dh P(h) \ln \left[\frac{\cos(h)^2}{(2 - \cos(h)^2) (\cos k + \sin(h)^2)^2 }\right] \;.
\end{equation}

For the zero momentum sector, instead two states can contribute to the trace, i.e. the vacuum $\ket{\bf 0}$ and the zero-momentum magnon $\ket{k = 0}$. The trace in this sector can then be rewritten as
\begin{equation}
    \Tr_{k=0}[\VV^{(1)} \VV^{(2)}_{1} \ldots \VV^{(1)} \VV^{(2)}_{L}] = (   \cos J)^{t L}
\Tr[M_1\ldots M_L]
\end{equation}
where the matrices $M_j = M(h_j)$ and $M(h)$ take the form
\begin{equation}
M(h) = 
\left(
\begin{array}{cc}
 2 \cos (t \theta_h ) & -2 \imath \sqrt{t} \cos (\alpha_h ) \sin (\theta_h  t) \\
 2 \tan (J) \sqrt{t} \cos (\alpha_h ) \sin (\theta_h  t) & 2 \imath \tan (J) \left(t \cos (\theta_h  t) \cos(\alpha_h )^2+\cot (\theta_h ) \sin (\theta_h  t) \sin(\alpha_h )^2\right) \\
\end{array}
\right)    
\end{equation}
By computing the two Lyapunov exponents $\eta_0, \eta_1$ associated with the sequence of random matrices $M_j$ (see the method explained in Sec.~\ref{sec:numerics})
\begin{equation}
    \Tr[M_1\ldots M_L] \longrightarrow A_0 e^{\eta_0 L} + A_1 e^{\eta_1 L}
\end{equation}
we have the approximation 
\begin{equation}
    \lambda_0(k=0) \sim 2 t \ln |\cos(J)| + 2 \eta_0  \;, \quad 
    \lambda_1(k=0) \sim 2 t \ln |\cos(J)| + 2 \eta_1  \;.
\end{equation}

\section{$\langle \mathcal{K}^n(t)\rangle $ in the MBL phase} \label{app:mbl_hpsff}

In this Appendix we analyze $\langle \mathcal{K}(t)^\alpha \rangle$ for the LIOM model \eqref{eq:liom} with 2-body nearest-neighbour terms for integer $\alpha$. We map $\langle \mathcal{K}(t)^\alpha \rangle$ to the partition function of stacked spin chains with 2-body interactions, which can be written in terms of a transfer matrix\cite{znidaric18}. We numerically diagonalize the transfer matrix constructed from the LIOM and show that the results are qualitatively compatible with the numerical results from the RP and KIM model in MBL regime.

\begin{figure}[t!]
	\includegraphics[width=0.4\columnwidth]{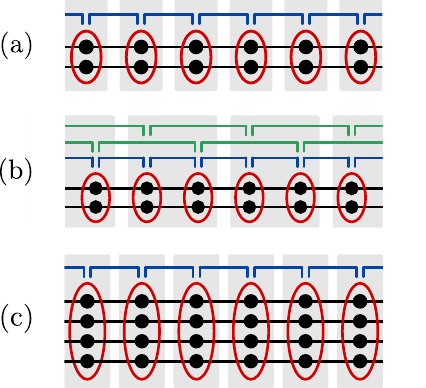}
	\caption{(a) Representation of 
		$\langle \mathcal{K}(t)\rangle$ with nearest-neighbour 2-body interactions, which is mapped to the partition function of a pair of spin chains with 1-body (red) and nearest-neighbour (blue) 2-body interactions.
		The grey regions illustrate the Hilbert space associated with the transfer matrix for $\langle \mathcal{K}(t)\rangle$, which has $4$ d.o.f.
		(b) Representation of 
		$\langle \mathcal{K}(t)\rangle$ with up to next-to-nearest-neighbour 2-body interactions, which is mapped to spin chains with additional next-to-nearest-neighbour (green) 2-body interactions. The associated  Hilbert space has dimension $16$. (c) Representation of $\langle \mathcal{K}(t)^2 \rangle$ with nearest-neighbour 2-body interaction, which is mapped to four spin chains with 1-body (red) and nearest-neighbour terms (blue). Again, the Hilbert space dimension is $16$.
	} \label{fig:liom_tmat}
\end{figure}

It is instructive to construct the transfer matrix for $\langle \mathcal{K}(t) \rangle$ for \eqref{eq:liom} with nearest-neighbour 2-body terms, and then generalize the procedure for general 2-body terms and hpSFF. Before averaging, the argument of the (1st point) SFF is
\begin{align}
\mathcal{K}(t) = 
&\sum_{ \{ \mathbf{m}, \mathbf{n} \}}  \prod_{k=1}^{L}   \exp \Bigg[ \imath t   J^{(1)}_k  \left(  m_k - n_k \right) 
 +  \imath t   J^{(2)}_{k,k+1}  \left(  m_k m_{k+1} - n_k n_{k+1}\right)  \Bigg] \; ,
\end{align}
where $\mathbf{m} = (m_1, m_2, \dots, m_L)$, $m_k= \pm 1 $, and the first sum is over all possible values of $\mathbf{m}$ and $\mathbf{n}$. $J^{(1)}_k$ and $ J^{(2)}_{k,k+1}$ are distributed according to \eqref{eq:j1j2}. The ensemble average gives
\begin{align} \label{eq:kavgising}
\langle \mathcal{K} (t) \rangle =  
&\sum_{ \{ \mathbf{m}, \mathbf{n} \}} 
\prod_{k=1}^L
\exp \Bigg[  - \frac{1}{2}t^2  J_1^2  \left(  m_k - n_k \right)^2
-\frac{1}{2}t^2  J_2^2 \left(  m_k m_{k+1} - n_k n_{k+1}\right)^2  \Bigg] \; .
\end{align}
This is the partition function of a stack of two spin chains whose state is specified by $\mathbf{m}$ and $\mathbf{n}$, see Fig.~\ref{fig:liom_tmat}a. Consider the basis, $(m_k, n_k)$ with $m_k, n_k = \pm 1$.  Eq.~\ref{eq:kavgising} can then be re-written using a transfer matrix in terms this basis as
\begin{equation}
T = 
\begin{bmatrix}
1 & h_1 h_2 & h_1 h_2 & 1
\\
h_1 h_2 & h_1^2  & h_1^2  & h_1 h_2
\\
h_1 h_2 & h_1^2  & h_1^2  & h_1 h_2
\\
1 & h_1 h_2 & h_1 h_2 & 1
\end{bmatrix} \; ,
\end{equation}
where $h_1 =  \exp(-   t^2  J_1^2 )$ and  $h_2 =  \exp(- t^2  J_2^2 )$, and 
\begin{equation}
\langle \mathcal{K}(t) \rangle = \mathrm{Tr} \left( T^L \right)   \;,
\end{equation}
for the periodic boundary condition (the case of open boundary condition can also be evaluated).
The diagonalization of $T$ gives two eigenvalues of 0 with eigenvectors $(-1,0,0,1)^T$ and $(0,-1,1,0)^T$. The non-vanishing eigenvalues are 
\begin{align}
E_{\pm} &= 
1+ e^{-2 J_1^2 t^2}
\pm \sqrt{4 e^{-2 t^2 \left(J_1^2+2 J_2^2\right) }
	+e^{-4 J_1^2 t^2}-2 e^{-2 J_1^2 t^2}+1}
\end{align}
and we have in this case
\begin{equation}
 F_t(\alpha = 1) = \ln E_+
\end{equation}
while $E_-$ corresponds to the second Lyapunov exponent.    
As a consistency check, in the uncoupled regime where $J_2 = 0$, we have only a single non-degenerate exponent,
\begin{align}
&E_+ = 
2+ 2 e^{-2 J_1^2 t^2}
\\
& E_-= 0 \; .
\end{align}
With periodic boundary condition, $\langle\mathcal{K} (t) \rangle |_{J_2 = 0}= (2+2 e^{-2 J_1^2 t^2})^L \to 2^L$ at large $t$ as expected. 

The evaluation of $\langle \mathcal{K}(t) \rangle$ can be generalized to LIOM \eqref{eq:liom} with general (not just nearest-neighbour) 2-body terms. We take the variance of 2-body coupling between spins separated by $r$ sites to be $\langle  (J_{i,i+r}^{(2)})^2\rangle = J_2^2 \, e^{-2(r-1)/ \xi} \equiv J_{2,r}^2$.
Using the same approach, the ensemble average becomes 
\begin{align} 
\langle \mathcal{K}(t) \rangle =  
&\sum_{ \{ \mathbf{m}, \mathbf{n} \}} 
\prod_{k=1}^L
\exp \Bigg[  - \frac{1}{2}t^2  J_1^2  \left(  m_k - n_k \right)^2
- \sum_{r=1}^{r_{\mathrm{max}} } \frac{1}{2}t^2  J_{2,r}^2 \left(  m_k m_{k+r} - n_k n_{k+r}\right)^2  \Bigg] \; . \label{eq:2bod_general_r}
\end{align}
This is the partition function of a stack of two spin chains with 2-body interactions up to a distance of $r_{\mathrm{max}}$. Consequently, the Hilbert space associated with the transfer matrix is a tensor product of $r_{\mathrm{max}}$ copies of on-site Hilbert spaces, and contains degrees of freedom labelled by $(m_k,n_k,\dots,m_{k+ r_{\mathrm{max}-1}}, ,n_{k+ r_{\mathrm{max}-1}})$, where $n_k, m_k\dots =\pm 1$. 
The cases of $r_{\mathrm{max}}=1$ and $2$ are illustrated in Fig.~\ref{fig:liom_tmat} a and b. 
The resulting transfer matrix has $4^{r_{\rm max}}$ eigenvalues: a genuine MBL phase has an infinite number of non-trivial Lyapunov exponents which are recovered in the limit $r_{\rm max} \to \infty$.

\begin{figure}[t!]
	\includegraphics[width=0.4\columnwidth]{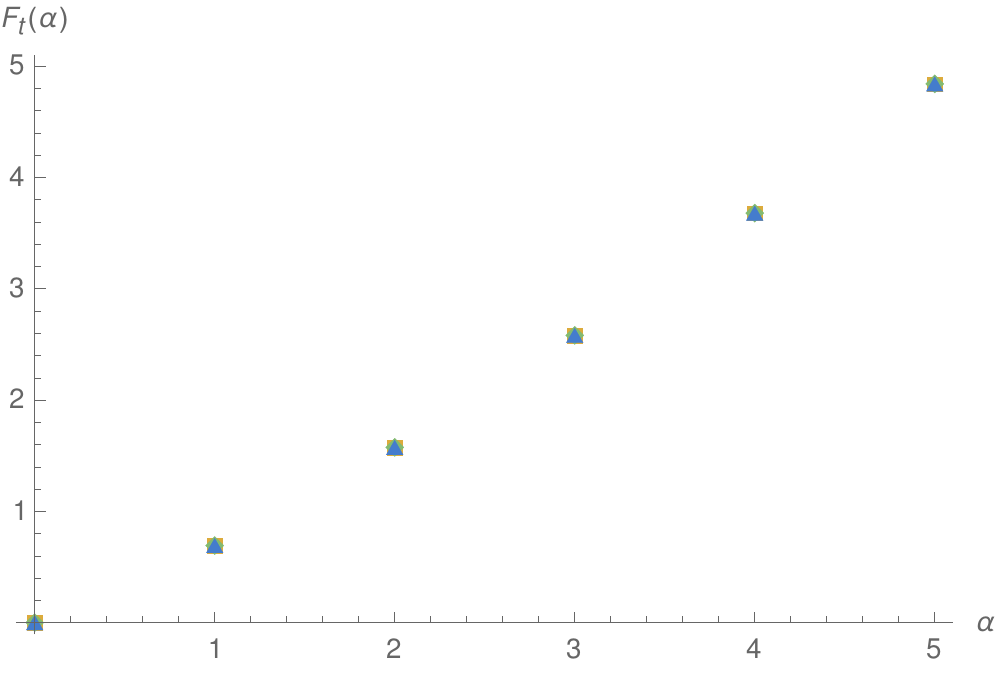}
	\caption{$F_t(\alpha)$ vs integer $\alpha$ for LIOM with 2-body nearest neighbour terms at $J_1 = J_2 = 1$ and $t=1,10,100,1000$, represented by different symbols. Other finite values of the ratio $J_1/ J_2$ give qualitatively similar behaviour for $F_t(\alpha)$.
	} \label{fig:liom_falpha_Jtwo1}
\end{figure}

We can further generalize this approach to the evaluation of $\langle K^n(t) \rangle$ with integer exponent $n$ and with only 2-body nearest-neighbour terms. In this case we have
\begin{align} \label{eq:kavgisinggen}
& \langle \mathcal{K}^n(t) \rangle = 
\sum_{\mathbf{m}^{(1)}, \mathbf{n}^{(1)}, \mathbf{m}^{(2)}, \mathbf{n}^{(2)}\dots } 
\prod_{k=1}^L
\exp \Bigg\{  - \frac{1}{2}t^2  J_{1}^2   \left[\sum_{i=1}^{n} \left(m^{(i)}_k - n^{(i)}_k \right)\right]^2
-\frac{1}{2}t^2  J_{2}^2  \left[\sum_{i=1}^{n}  \left( m^{(i)}_k m^{(i)}_{k+1} - n^{(i)}_k n^{(i)}_{k+1}\right)\right]^2  \Bigg\} \; ,
\end{align} 
which  is the partition function of $2n$ copies of spin chains with 2-body nearest-neighbour interaction, as illustrated in Fig.~\ref{fig:liom_tmat} c, so that the transfer matrix Hilbert space size grows as $4^n$.
We numerically diagonalize the transfer matrix, and plot the value of $F_t(\alpha)$ in Fig.~\ref{fig:liom_falpha_Jtwo1} for integer $\alpha$ up to $\alpha=5$. Although this approach does not allow analytical continuation of $\langle \mathcal{K}^n(t) \rangle$, we see that the form of $F_t(n)$ is compatible with the expectation that $ \lambda_>  = F'_t(\alpha=0)$ is finite, as discussed in Sec.~\ref{sec:mbl_phase}. 

\section{Gaps in Lyapunov spectrum}\label{app:lyap_gap}
\begin{figure}[t!]
		\includegraphics[width=0.5\columnwidth]{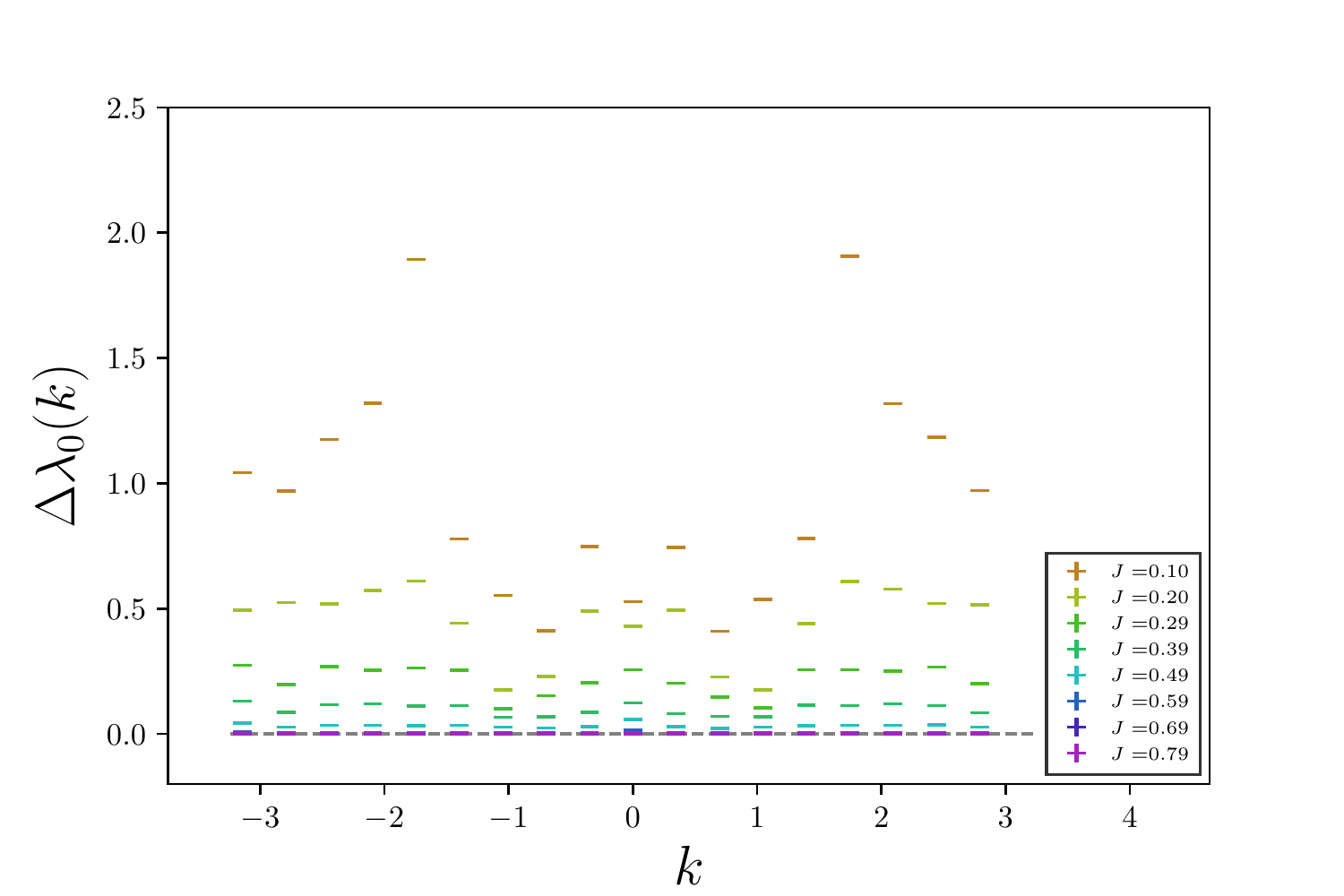}
		\caption{
		$\Delta \lambda_0(k)$ vs $k$ for KIM at $t=18$. Deep in the chaotic phase, $\Delta \lambda_0(k)$ are small for all $k$
		}  \label{fig:kim_gap}
	\end{figure}

\begin{figure}[t!]
		\includegraphics[width=0.5\columnwidth]{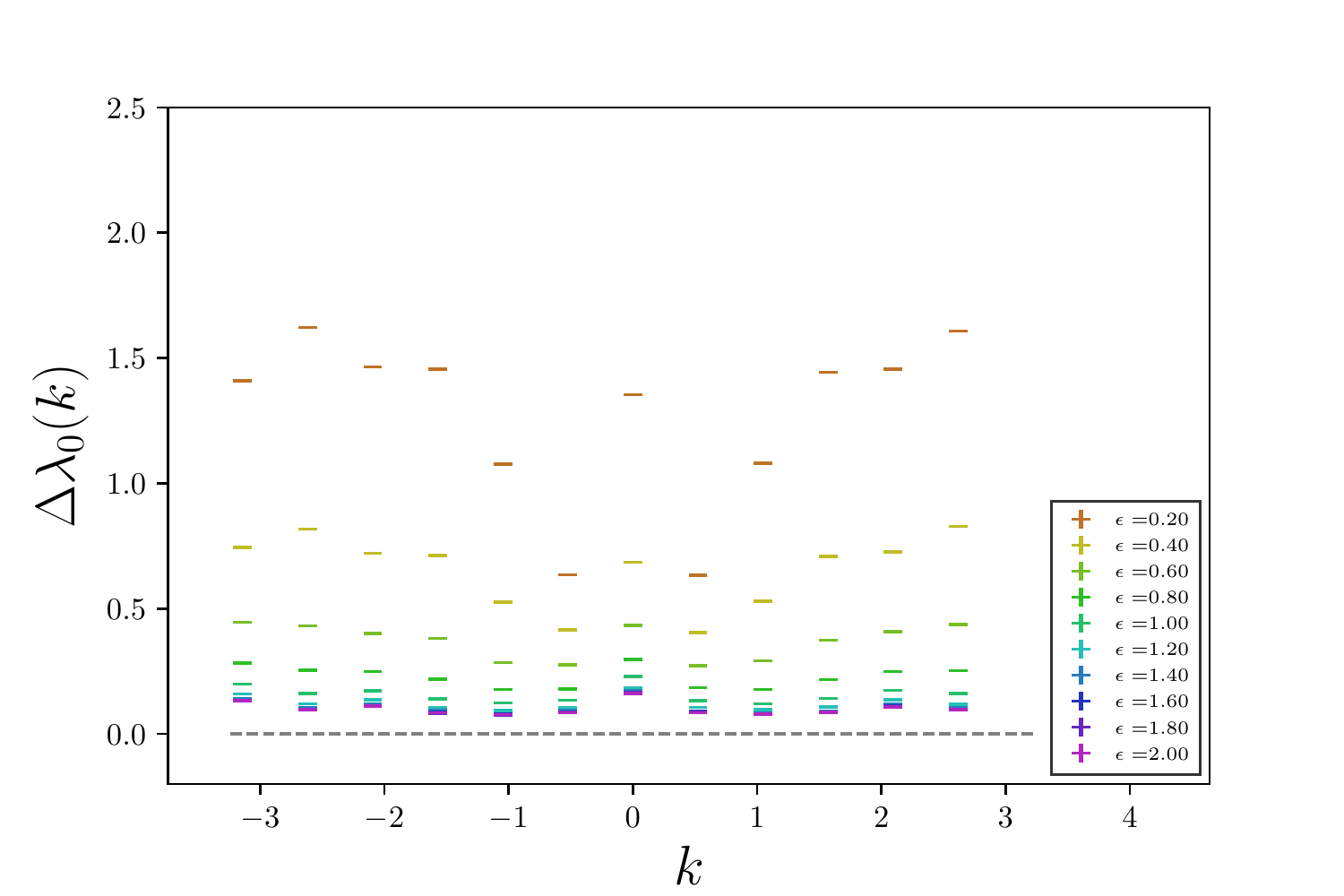}
		\caption{
		$\Delta \lambda_0(k)$ vs $k$ for  RPM at $t=12$. Deep in the chaotic phase $\Delta \lambda_0(k)$ remains gapped for all $k$.
		} \label{fig:rpm_gap}
	\end{figure}

In Fig.~\ref{fig:kim_gap} and \ref{fig:rpm_gap}, we show $\Delta \lambda_a(k)\equiv \lambda_a(k) -\lambda_{a+1}(k)$ with $a=0$ computed for the KIM and RPM. In particular, in the chaotic phase of the KIM,  the gap in the Lyapunov spectrum is small. This supports the expectation that, in each time-momentum sector, there are two vanishing Lyapunov exponents contributing to $\langle K(t) \rangle \sim \sum_{k,a} e^{\lambda_a(k) L}$ at long times. In the chaotic phase of the RPM, the corresponding computation suggests that $\Delta \lambda_a(k)$ remains gapped.

\end{document}